\documentclass[final,1p,times]{elsarticle}
\usepackage{amsmath,amssymb,amsfonts}
\usepackage{algorithmic}
\usepackage{graphicx,color}
\usepackage{textcomp}
\usepackage[dvipsnames]{xcolor}
\usepackage{hyperref}
\usepackage{algorithm,algorithmic}
\usepackage{multirow}
\usepackage{placeins}
\usepackage{cleveref}
\usepackage{enumitem,comment}

\usepackage{fancyhdr}
\usepackage{hyperref}

\makeatletter

\newcommand{\ManuscriptHeader}{%
  \footnotesize
  Accepted manuscript. Formal published version:
  \href{https://doi.org/10.1016/j.ultras.2026.108090}{doi.org/10.1016/j.ultras.2026.108090}%
}

\newcommand{\ManuscriptFooter}{%
  \tiny
  \textcopyright\ 2026. This manuscript version is made available under the
  \href{https://creativecommons.org/licenses/by-nc-nd/4.0/}{CC-BY-NC-ND 4.0 license}%
}

\fancypagestyle{plain}{%
  \fancyhf{}
  \fancyhead[R]{\ManuscriptHeader}
  \fancyfoot[C]{\ManuscriptFooter}

}

\newcommand{\unc}[1]{\text{\tiny{#1}}}
\newcommand{\tightpm}{\ensuremath{{\scriptstyle\pm}}}
\newcommand{\eg}{e.g., }
\newcommand{\ie}{i.e., }

\begin{document}

\begin{frontmatter}

\title{Image-Based Metrics in Ultrasound for\\ Estimation of Global Speed of Sound}
\tnotetext[thanksref]{Funding was provided by the Centre for Interdisciplinary Mathematics and the Medtech Science and Innovation Centre at Uppsala University, Sweden.}

\author[label1]{Roman Denkin\corref{cor1}}
\ead{roman.denkin@it.uu.se}

\author[label1]{Orcun Goksel}

\cortext[cor1]{Corresponding author.}

\affiliation[label1]{organization={Department of Information Technology, Uppsala University},
             city={Uppsala},
             country={Sweden}}

\begin{abstract}
Accurate speed-of-sound (SoS) estimation is crucial for ultrasound image formation, yet conventional systems often rely on an assumed value for imaging. 
We propose to leverage conventional image analysis techniques and metrics as a novel and simple approach to estimate tissue SoS. 
We study eleven metrics in three categories for assessing image quality, image similarity and multi-frame variation, by testing them in numerical simulations and phantom experiments, as well as testing in an in vivo scenario.
Among single-frame image quality metrics, conventional Focus and a proposed metric variation on Tenengrad present satisfactory accuracy (5-8\,m/s on phantoms), but only when the metrics are applied after compounding multiple frames. 
Differential image comparison metrics were more successful overall with errors consistently under 8\,m/s even applied on a single pair of frames. 
Mutual information and correlation metrics were found to be robust in processing relatively small image patches, making them suitable for focal estimation. 
We present an in vivo study on breast density classification based on SoS, to showcase clinical applicability.
The studied metrics do not require access to raw channel data as they can operate on post-beamformed and/or B-mode data.
These image-based methods offer a computationally efficient and data-accessible alternative to existing physics- and model-based approaches for SoS estimation.
\end{abstract}


\end{frontmatter}

\section{Introduction}
Ultrasound imaging systems collect data in the time domain that require projection into the spatial domain to create images of internal tissue structures. 
This process, known as beamforming, forms the foundation for many ultrasound-based downstream imaging techniques. 
The conversion between the domains of time and space necessitates knowledge of the conversion factor, specifically the speed-of-sound (SoS) in the imaged sample.
Typically, an assumed region-specific SoS (\eg 1540\,m/s) is used. 
However, the actual SoS values can vary significantly across the population and between different tissue types~\cite{goss1980compilation}. 
The oversimplification to an assumed constant value leads to various imaging artifacts (localization and smoothing) and reduced image resolution.
To address these limitations, multiple methods have been developed to estimate tissue SoS based on various physical models of sound propagation. 
Second-order polynomials were fitted to echo profiles in~\cite{flax1988phaseaberration,anderson1998direct}. 
Speckle shifts between frames obtained from multiple transmit events were minimized in~\cite{bezek2023analytical}.

Multiple methods employ optimization of some quantity calculated from images beamformed using different assumed SoS values.
A focus quality metric based on frequency spectrum analysis is maximized in~\cite{napolitano2006sound}.
In~\cite{shin2010estimation}, a correlation-based metric is maximized, which measures restoration quality of deconvolutions of RF ultrasound data with point-spread functions (PSFs) simulated at different sound speeds. 
Minimum average phase variance optimization of RF channel data after applying the dynamic focusing delay patterns is used in~\cite{yoon2011vitro}.
The above method was extended in~\cite{park2011mean} with the optimization of minimum average sum of the absolute difference of raw radio-frequency (RF) data during receive beamforming.
Ultrasound speckle shape, quantified by full-width-at-half-maximum of autocovariance function, is maximized in~\cite{qu2012average}.
Autocorrelation of ultrasound images was used to characterize the size of speckle in ultrasound speckle pattern under the assumption that better assumed beamforming SoS produces sharper well-defined speckles. 
In~\cite{anderson2000impact}, these methods are used to solve the inverse problem of quantifying errors caused by incorrect SoS assumption. 
Autocovariance-based methods have been used to find optimal beamforming SoS in~\cite{qu2012average}.
In~\cite{hasegawa2019initial}, coherence factor is maximized to search for global SoS.
Signal coherence between different transmit (Tx) and receive (Rx) paths was optimized in plane-wave imaging in~\cite{shen2020dmas}.
In~\cite{perrot2021das}, a phase-based quality metric that analyzes phase uniformity along hyperbolic signal paths during delay-and-sum beamforming is maximized. 
In~\cite{he2009sound}, spectrum energy of lateral PSF estimation from raw RF images is maximized to find the optimal imaging SoS, closely related to speckle brightness as quality factor in~\cite{nock1989phase}.
Fuzzy-logic-based algorithm is used to combine multiple image properties into a single SoS estimator in~\cite{he2017sound}, incorporating mean value, energy and contrast of an image.
Combined sharpness and brightness of a focal area is used to estimate SoS in~\cite{benjamin2018surgery}.
\Cref{tab:sos_methods_comparison} provides a non-exhaustive list of global SoS estimation methods in the literature, with their data and acquisition requirements. 
Most earlier methods utilize raw RF channel data, whereas methods operating on beamformed or B-mode data offer broader accessibility, also motivating our study into image-based metrics.

\begin{table*}[t]
\centering
\small
\setlength{\tabcolsep}{4pt}
\renewcommand{\arraystretch}{1.15}
\caption{Overview of existing global (homogeneous-equivalent) SoS estimation methods, summarizing their approach, input data type, 
whether raw channel access is needed, and the number of transmit acquisition events required. 
}
\label{tab:sos_methods_comparison}
\resizebox{\textwidth}{!}{
\begin{tabular}{|c c c c c|}
\hline
Paper & Method & Data type & Channel access & Transmissions\\
\hline
Anderson (1998)~\cite{anderson1998direct} &
Direct/Parabolic fit to echo signal &
Raw RF channel &
Yes &
Single  \\
\hline
Napolitano (2006)~\cite{napolitano2006sound} &
Lateral spatial frequency spectrum optimization &
Raw RF channel &
Yes &
Single\\
\hline
He (2009)~\cite{he2009sound} &
Lateral PSF optimization &
Beamformed RF &
No &
Multiple trials \\
\hline
Shin (2010)~\cite{shin2010estimation} &
Deconvolution with model PSFs to minimize lateral autocorrelation &
Beamformed RF &
No &
Single \\
\hline
Park (2011)~\cite{park2011mean} &
Minimization of average sum of the absolute difference in RF data&
Raw RF channel &
Yes &
Single \\
\hline
Yoon (2011)~\cite{yoon2011vitro} &
Minimization of average phase variance in RF data&
Raw RF channel &
Yes &
Single \\
\hline
Qu (2012)~\cite{qu2012average} &
Minimization of average normalized autocovariance function &
Raw RF channel &
Yes &
Single \\
\hline
He (2017)~\cite{he2017sound} &
Fuzzy logic aggregation of energy, contrast, and mean intensity &
Beamformed I/Q &
No &
Multiple trials \\
\hline
Benjamin (2018)~\cite{benjamin2018surgery} &
Combination of brightness, FFT spectrum intensity, and gradient direction &
B-Mode &
No &
Multiple trials \\
\hline
Hasegawa (2019)~\cite{hasegawa2019initial} &
Maximization of coherence factor &
Raw RF channel &
Yes &
Single \\
\hline
Shen (2020)~\cite{shen2020dmas} &
Maximization of DMAS coherent factor &
Raw RF channel &
Yes &
21 plane waves \\
\hline
Perrot (2021)~\cite{perrot2021das} &
Maximization of phase-quality metric &
I/Q channel &
Yes &
Single \\
\hline
Bezek (2023)~\cite{bezek2023analytical} &
Analytical SoS updates from displacement between different TX events &
Beamformed RF &
No &
2+ DW \\
\hline
Xiao (2024)~\cite{xiao2024realtime} &
Optimization of coefficient of variation/autocorrelation across different TX events &
Raw RF channel &
Yes &
11/31 plane waves \\
\hline
\end{tabular}}
\end{table*}

Several methods above use the term \emph{mean} SoS when referring to the optimal homogeneous-equivalent single SoS value for an imaged region.
Note that this can be somewhat misleading since the mean (average) value of a heterogeneous SoS distribution would not optimize the image appearance, but instead the SoS value that statistically minimizes the beamforming time-delay errors (TDE) need to be used~\cite{bezek2023analytical}.
Conversely, any method that optimizes the beamformed image quality in a region will find that same TDE-minimizing SoS value.
For such single, homogeneous-equivalent, beamforming SoS value, we prefer the term \emph{global SoS} as in~\cite{bezek2022global,strohm2022ius} to help contrast it with its locally-resolved imaging counterpart called typically as local SoS.

Note that the beamforming image quality from the choice of global SoS affect several downstream US imaging techniques and tasks.
For instance, local SoS reconstruction methods may rely on the initial global SoS assumption~\cite{bezek2023analytical} and shear-wave elastography measurements were shown to vary largely given beamforming SoS choices~\cite{chintada2021phase}.
Global SoS estimation is also practically very relevant as beamforming on most ultrasound systems inherently requires a single SoS assumption and cannot directly incorporate spatially-varying local SoS maps. 
Furthermore, in morphologically uniform tissue regions, global SoS can be used for tissue quantification as well, such as for muscles~\cite{xiao2024realtime} and the breast density application shown in our work.

In this study, we propose a novel approach that leverages image-analysis techniques from photography and other imaging applications, \eg for adjusting focusing, to assess US image quality as well as to quantify US image similarity for varying beamforming SoS values. 
Note that focus quality refers specifically to minimizing the point spread function (PSF) spatially, whereas the image-based metrics studied herein may also capture other image characteristics, e.g., based on global intensity distribution.
These image processing techniques may provide a robust alternative for determining accurate SoS values, both for US signal-to-image reconstruction as well as a diagnostic imaging biomarker. 
By studying image processing methods, we aim to develop an SoS estimation approach that reduces reliance on physics-based models, which may depend on sequence choices and machine specifics, cumbersome to compute, complex to formalize, and hence potentially error-prone and inaccurate.
We herein omit metrics based on physical US properties, \eg methods based on PSF analysis, due to their dependency on transmit sequence choices and propagated medium.
Purely image-based approach are analyzed in this work, as being more generalizable and wider applicable thanks to easier access to beamformed data on most US systems.
We study several methods comparatively, in particular exploring the image similarity metrics, which we show as promising for SoS estimation.

\section{Methods}
We study multiple image analysis metrics in three distinct categories:
\textit{Image quality} metrics evaluate focusing-related image features and can be computed on a single image frame or on compounding of multiple frames for increased signal-to-noise ratio. 
These methods assume that the images should be maximally-focused (least-blurred) when the beamforming SoS is the global (TDE-minimizing) SoS value.
\textit{Image similarity} metrics compare two frames covering the same imaging area but obtained with different Tx events viewing them from different directions.
This utilizes the fact that the Tx-path differences to an image pixel would change (either constantly across the image or a function of location, given the sequence), which will co-align the pixels at the global SoS value.
\textit{Multi-frame statistic} metrics extend the similarity of frame-pair alignment to multi-frame co-occurrence where pixel intensity values across multiple frames are considered minimizing their variation in some form.

To determine the optimal tissue SoS, we evaluate these metrics across a range of SoS values and define the ideal SoS value that optimizes the metric. 
Although various optimization methods could be employed to accelerate such search, as we aim to assess metric optimality independent of optimization approaches, in this work we employ a grid-search approach computing metric values on a fine resolution across a predefined range of SoS settings. 
Based on the optimality characteristics of individual metrics, in practical settings one can implement them on coarser evaluation grids with interpolation in-between, and/or by using iterative approaches based on optimization.

\begin{figure}
\centerline{\includegraphics[width=0.9\textwidth,trim={3cm 2cm 5cm 0cm},clip]{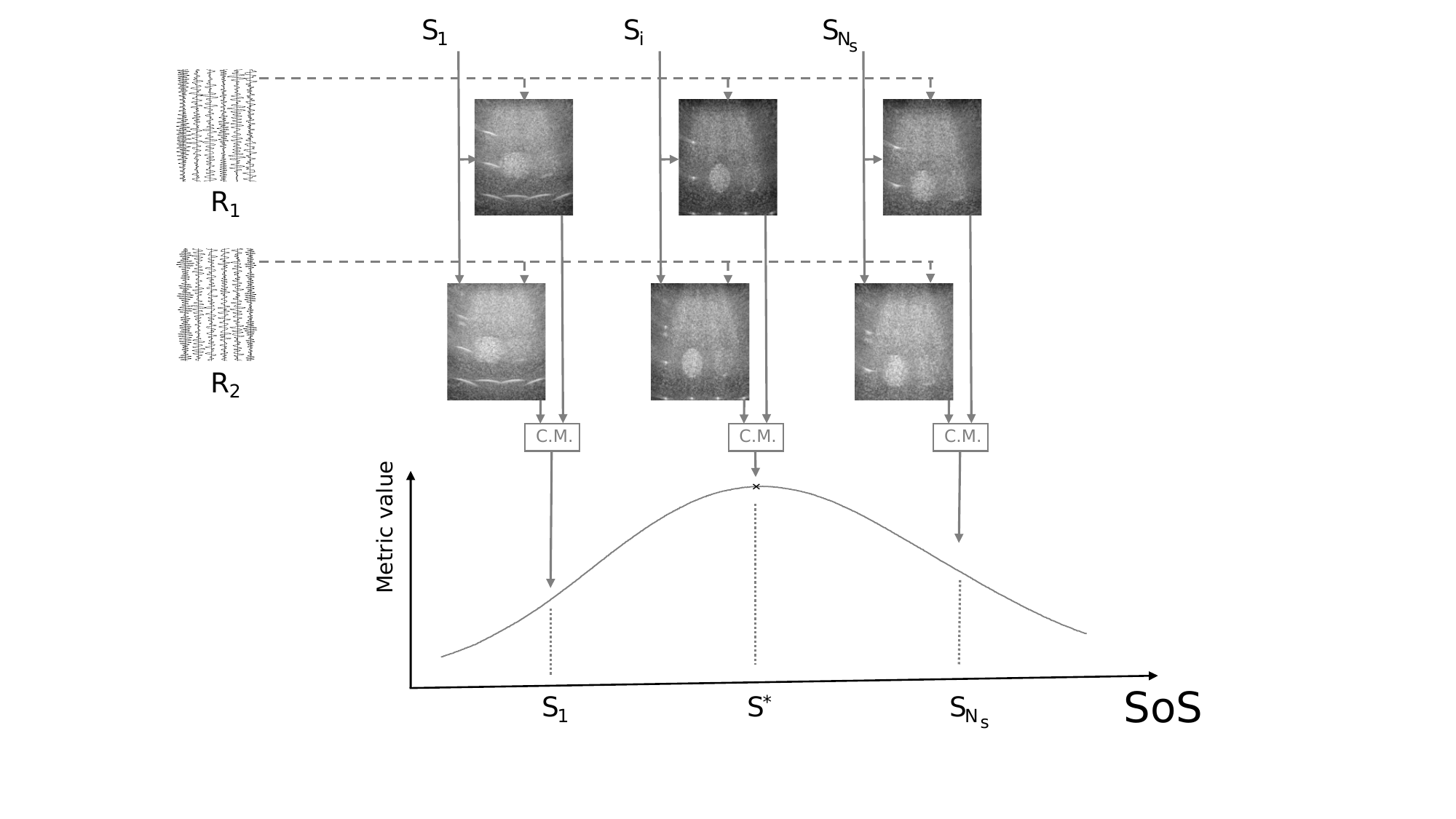}}
\caption{Pipeline overview for optimizing a comparison metric (C.M.) between two acquisitions ($R_1$ and $R_2$) using trial-and-error of different SoS values. 
Note that image quality and multi-frame statistical metrics work similarly, but instead with one or several acquisitions, respectively.}
\label{pipeline_fig}
\end{figure}

Our processing pipeline for finding the optimal SoS employs US RF echo data $R\in \mathcal{R}^{n_\mathrm{r}\times n_\mathrm{k}}$ of $n_\mathrm{r}$ receiving elements for $n_\mathrm{k}$ time samples from a Tx event. 
We may also obtain multiple acquisitions
$\{R_1, \ldots, R_{n_\mathrm{t}}\}$ from $n_\mathrm{t}$ Tx events with different transmit sequences to get multiple views of the same tissue for image comparison or to increase signal-to-noise ratio of images by compounding. 
RF echo signals are beamformed into spatial RF images $I^{n_t}(s_i)$ using a range of $n_\mathrm{s}$ assumed SoS values $\{s_1, \ldots, s_{n_\mathrm{s}}\}$.
Image comparison metrics are calculated based on pairs of RF images, while image quality metrics are applied on B-mode images (\ie after envelope detection and log compression).
An overview for the image comparison setting is illustrated in \Cref{pipeline_fig}.
The optimal global SoS is then determined as $s^\star = \arg\max_{s_i} m(\{I^{n_t}(s_i)\})$, where $m(\cdot)$ denotes the chosen metric evaluated on the beamformed image(s) at each trial SoS value $s_i$.

\subsection{Image Quality Metrics}
Let $I \in \mathbb{R}^{n_x \times n_z}$ represent a beamformed ultrasound envelope image, where $I$ is a matrix of pixel intensities on an $x$-$z$ grid of size $n_x \times n_z$. 
We study the following metrics.

\subsubsection{Focus}
This metric is based on the observation~\cite{de2013sharpness} for natural images that certain frequency components in the Fourier domain are maximized when optimal image sharpness is achieved. 
We implement this for US imaging as
\begin{equation}
\frac{\sum_{(f_x, f_z) \in F} |\mathrm{FFT}(I)_{(f_x, f_z)}|^2}{\sum |\mathrm{FFT}(I)|^2},
\end{equation}
where $f_x$ and $f_z$ are the spatial frequencies selected within a band $F$ that contain frequency magnitudes between $f_1$ and $f_2$, \ie satisfying the criterion $f_1 \leq \sqrt{f_x^2 + f_z^2} \leq f_2$ which represents a ring in the Fourier domain. 

\subsubsection{Entropy}
This leverages information theory principles to quantify image complexity based on the randomness of pixel intensity distributions, following the approach described in~\cite{sparavigna2019entropy}, with the metric defined on an intensity histogram as:
\begin{equation}
-\sum_{i=1}^{n_\mathrm{b}} p(i) \log_2 p(i),
\end{equation}
where $p(i)$ represents the probability of pixel intensities falling into histogram bin $i$, and $n_\mathrm{b}$ is the number of bins used for the histogram discretization. 
By assessing the distribution of pixel values, this metric evaluates the image's information content. 
High entropy suggests a potentially accurate SoS setting that uncovers details (\ie more information) in $I(x,z)$, whereas low entropy indicates an information-poor image.

\subsubsection{Tenengrad}
This metric is proposed in~\cite{krotkov1988focusing} and is based on the conventional understanding that a better focused (sharper) image would have larger gradient magnitudes, \ie
\begin{equation}  \label{eq:tenengrad}
\sum_{x,z} T(x,z) = \sum_{x,z} \sqrt{g_x^2(x,z) + g_z^2(x,z)}\,,
\end{equation}
where $g_x(x,z)$ and $g_z(x,z)$ are image gradients along the lateral and axial directions, respectively.

\subsubsection{ANACVF}
Average normalized autocovariance function (ANACVF) is proposed in~\cite{qu2012average}, defined as:
\begin{equation}
\frac{1}{n_\mathrm{u}n_\mathrm{r}}\sum_{(x,z) \in U}\sum_{(u,v) \in \mathrm{ROI}}\frac{[I(u+x,v+z)-\overline{I}][I(u,v)-\overline{I}]}{\sigma^2},
\end{equation}
where ROI is an US image region of interest on which to compute this metric, $U$ represents a range of shift values in lateral and axial directions, $\sigma^2$ is the intensity variance within ROI, and $n_\mathrm{u}$ and $n_\mathrm{r}$ are the set sizes of $U$ and ROI respectively.

\subsubsection{ST-Ten}
We propose an adaptation of the Tenengrad metric for US imaging, which consists of three sequential steps designed to suppress US-characteristic noise while enhancing sensitivity to SoS-related defocusing artifacts. 
We first smooth the image while preserving large structural features by using Gaussian filtering
\begin{equation}
I'(x,z) = \frac{1}{2\pi\sigma^2}e^{-(x^2+z^2)/(2\sigma^2)} * I(x,z)
\end{equation}
and compute the directional gradients of the smoothed image as
$g_x = \partial I'/\partial x$ and 
$g_z = \partial I'/\partial z$.
ST-Ten is defined similar to Tenengrad but by only considering the gradient magnitudes above a threshold $\tau$ as follows:
\begin{equation}
\sum_{x,z} T^2(x,z) = \sum_{x,z} \left(g_x^2(x,z) + g_z^2(x,z)\right), \ \ \forall_{T(x,z) \geq \tau}\,.
\end{equation}
The thresholding helps disregard many smaller gradients that would be less sensitive to defocusing from any incorrect SoS setting, which otherwise could overrun the total metric value.
Such thresholding thus allows the metric to focus on major structural edges that exhibit a clearer differentiation with SoS variation.

\subsection{Image Similarity Metrics}
These metrics are based on comparing two images, $I_1(x,z)$ and $I_2(x,z)$, each representing the same imaging field of view captured using differing acquisition sequences such that the sound waves to an imaged point travels different acoustic paths from transmission to echo receive; \eg plane waves with different transmission angles. 
When such acoustic paths differ, unless the correct SoS is used during beamforming to convert temporal signals into spatial images, the pixels of two images would not necessarily align.
By leveraging this information, image-similarity metrics can identify the tissue SoS value as the one that minimizes discrepancy between beamformed images.
Differing Tx paths are essential for these metrics, as identical paths would yield equally unfocused images that align for any SoS value; and it is the path-length differences that create SoS-dependent misalignment resolved only at the correct SoS.
To best of our knowledge, this is the first in-depth and comparative study of various image comparison metrics for their use to this end.

\subsubsection{Structural Similarity Index Metric (SSIM)}
This assesses image similarity based on a perceptual model considering texture, luminance, and contrast variations~\cite{wang2004ssim}, defined as:
\begin{equation}
\mathrm{SSIM}(I_1, I_2) = \left( \frac{2\mu_1\mu_2 + c_\mu}{\mu_1^2 + \mu_2^2 + c_\mu} \right) \cdot \left( \frac{2\sigma_{12} + c_\sigma}{\sigma_1^2 + \sigma_2^2 + c_\sigma} \right),
\end{equation}
where $\mu_1$ and $\mu_2$ represent the average pixel intensities, $\sigma_1^2$ and $\sigma_2^2$ denote their variances, and $\sigma_{12}$ is the covariance, within a small image patch that is scanned through the image as the metric is aggregated as a mean value between all patches. 
Constants $c_\mu$ and $c_\sigma$ ensure stability for small denominators for the corresponding term. 
The metric produces values between $-1$ and $1$, where the latter indicates perfect structural similarity.

\subsubsection{Mean Squared Error (MSE)}
This metric quantifies pixel-wise differences between two images, defined as:
\begin{equation}
\mathrm{MSE}(I_1, I_2) = \frac{1}{n_xn_z} \sum_{x=1}^{n_x} \sum_{z=1}^{n_z} (I_1(x,z) - I_2(x,z))^2\,.
\end{equation}
This metric aims for direct intensity comparison.
For consistency with the other introduced metrics, during SoS estimation we simply negate this metric, \ie -MSE, such that the metric value is positively correlated with image similarity and is maximized at zero for images that are identical.

\subsubsection{Peak Signal-to-Noise Ratio (PSNR)}
This metric evaluates image similarity by quantifying the ratio between the maximum possible signal and the noise power~\cite{huynhthu2008psnr}, defined as:
\begin{equation}
\mathrm{PSNR}(I_1, I_2) = 20 \cdot \log_{10}\left(\frac{\mathrm{MAX}(I_1\!\cup\! I_2)}{\sqrt{\mathrm{MSE}}}\right),
\end{equation}
where $\mathrm{MAX}(I_1\!\cup\! I_2)$ represents the maximum possible pixel value within both images.
Note that the denominator is conventional root mean square error (RMSE), used here for assessing noise. 
PSNR metric is hence a normalized (and log-compressed) version of RMSE, where the normalization makes the metric interpretable and the square-root operator compared to MSE reduces sensitivity to few pixels with large differences.

\subsubsection{Mutual Information (MI)}
This metric measures statistical dependency between intensity distributions of two images, defined as:
\begin{equation}
\mathrm{MI}(I_1,I_2) = \sum_{i}\sum_{j}p(i,j)\cdot \log\left(\frac{p(i,j)}{p(i)p(j)}\right),
\end{equation}
where $p(i,j)$ represents the joint probability distribution of intensity values, and $p(i)$ and $p(j)$ are the marginal probability distributions of $I_1$ and $I_2$ intensities, respectively, computed from a 2D joint-intensity histogram of two images with $n_\mathrm{b}^2$ bins. 
The metric thus reaches its maximum when intensity distributions exhibit strong statistical dependency, indicating alignment between the images, without assuming linear or any algebraic relation between image intensity correspondences.
Accordingly, this metric is commonly used also for cross-modality image registration.

\subsubsection{Correlation Coefficient (CC)}
This quantifies the linear relationship between pixel intensities, defined as:
\begin{equation}
\frac{\sum_{x,z} (I_1(x,z) - \bar{I}_1)(I_2(x,z) - \bar{I}_2)}{\sqrt{\sum_{x,z} (I_1(x,z) - \bar{I}_1)^2 \sum_{x,z} (I_2(x,z) - \bar{I}_2)^2}},
\end{equation}
where $\bar{I}_1$ and $\bar{I}_2$ denote the mean pixel values of respective images. 
The metric ranges between $-1$ and $1$, where the latter indicates perfect positive correlation suggesting optimal image alignment and hence a good beamforming SoS estimate.
It is commonly used for intra-modality image registration for providing robustness to any uncontrollable linear intensity variations.

\subsection{Multi-Frame Statistics: Coefficient of Variation (CV)}

This metric was proposed in~\cite{xiao2024realtime} based on minimizing a normalized form of pixel-wise variance across multiple frames, to evaluate the similarity between all frames collectively. 
The metric is defined for a set $\{I_i\}$ of $n_\mathrm{t}$ images, \ie $i\in\{1,2,..,n_\mathrm{t}\}$ as:
\begin{equation}
\mathrm{CV}(\{I_i\}) = \sum_{x,z}\frac{\sigma_\star(x,z)}{|\mu_\star|(x,z)},
\end{equation}
where $\sigma_\star(x,z)$ is the standard deviation and $|\mu_\star|(x,z)$ indicates the mean of the absolute pixel magnitudes at an image location $(x,z)$ across all the given $n_\mathrm{t}$ frames. 
This metric normalizes variance with average intensity to account for larger potential variances at higher magnitudes.

\section{Materials and Experiments}
\subsection{Setup}
For evaluation we first used simulations with RF data obtained using k-Wave simulation software~\cite{treeby2010kwave}.
We generated three homogeneous tissue phantoms with known ground-truth speeds of sound $\hat{s} = \{ 1400, 1500, 1600\}$\,m/s.
We modeled a linear transducer as used in the phantom experiments below, placed on a numerical domain of size 40$\times$55\,mm. 
Spatial and temporal simulation resolutions were set to be isotropic 75\,$\mu$m and 6.25\,ns, respectively. 
Tissue scatterers were simulated by slightly perturbing the medium density for a random 10\% of the simulation grid pixels.

For evaluation on real data, we imaged two phantoms using a UF-760AG ultrasound system (Fukuda Denshi, Japan) and a linear transducer with 128 elements, 300\,$\mu$m pitch, and 5\,MHz center frequency. 
The first, Phantom1, is CIRS 040GSE (Norfolk, VA, USA) with a declared SoS of 1540\,m/s.
The second, Phantom2, is a custom CIRS SoS phantom, where the imaging was conducted within its homogeneous background region with a declared SoS of 1509\,m/s. 
On each phantom, six acquisitions were collected by mechanically fixing the probe at different locations.

A diverging wave Tx sequence was employed with a Tx aperture of 31 elements and a virtual source (VS) 9\,mm behind the transducer surface, as in~\cite{schweizer2023robust}.
For all metrics, the entire imaging width of 38\,mm and an axial field-of-view of 32\,mm from $z=[8,40]$\,mm was considered to reduce the impact of near-field effects. 
In this imaging region, beamforming was performed for a grid of $n_\mathrm{r}\times n_\mathrm{k}$=256$\times$3072 RF samples using Delay-and-Sum (DAS) algorithm on echo data received on all elements. 
RF data caching and storage transfer performance define the delay between sequential VS sequence acquisitions of 37.5\,ms.

To assess image-quality metrics that require a single-frame, we used the B-mode images obtained with a single VS Tx at the center of the transducer; with this sequence and data referred hereafter as \emph{Single}.
For image-comparison metrics, we used the beamformed RF data from two VS Tx events separated by 3.6\,mm symmetrically around the center, called \emph{Dual}.
For assessing the multi-frame metric CV, we used the beamformed RF data from 17 Tx events, called hereafter \emph{Full}.
As the image-quality metrics can operate on arbitrary image input, we also tested such first group of methods on the Dual and Full acquisitions by compounding their beamformed RF frames before converting them to B-mode as input to image quality metrics.
B-mode images were produced via Hilbert transform in the temporal axis, log scaling, and grayscale mapping within a dynamic range of 60\,dB. 
For \emph{Dual} and \emph{Full} modes, beamformed RF frames were compounded first and then converted into B-mode images as described.

Metric sensitivity to motion and heterogeneous SoS has been tested on Phantom2 by moving the probe with a motorized motion stage in axial and lateral directions at controlled speeds as described in~\cite{schweizer2023robust}. 

\subsection{Implementation}
Several parameters such as the frequency bands and the Gaussian smoothing kernel size for ST-Ten were set empirically based on grid search using the compounded Dual data from the simulation with $\hat{s}=1500$\,m/s.
This yielded $f_1=0$, $f_2=0.1$, $tau=0.1$, and a Gauss kernel size of 5$\times$5.
For gradient-based metrics (Tenengrad and ST-Ten), a Sobel kernel size of 3 was used to calculate spatial intensity gradients~\cite{krotkov1988focusing}.
The following metrics were implemented using standard libraries with default hyper-parameters:
For SSIM and Entropy, \texttt{structural\_similarity} and \texttt{shannon\_entropy} functions from the scikit-image Python package were used with their default parameters.
For MI, \texttt{mutual\_info\_score} was used from the scikit-learn Python package by setting $n_\mathrm{b}=20$ bins.
For ANACVF metric, we used the settings proposed by its authors  in~\cite{qu2012average} with ROI as the entire image and the autocovariance shifts performed laterally up to 20 pixels with no axial shifts, \ie $z=0$ and $x\in\{-20,-19,..,19,20\}$. 

\begin{figure}
\centerline{\includegraphics[width=0.8\textwidth]{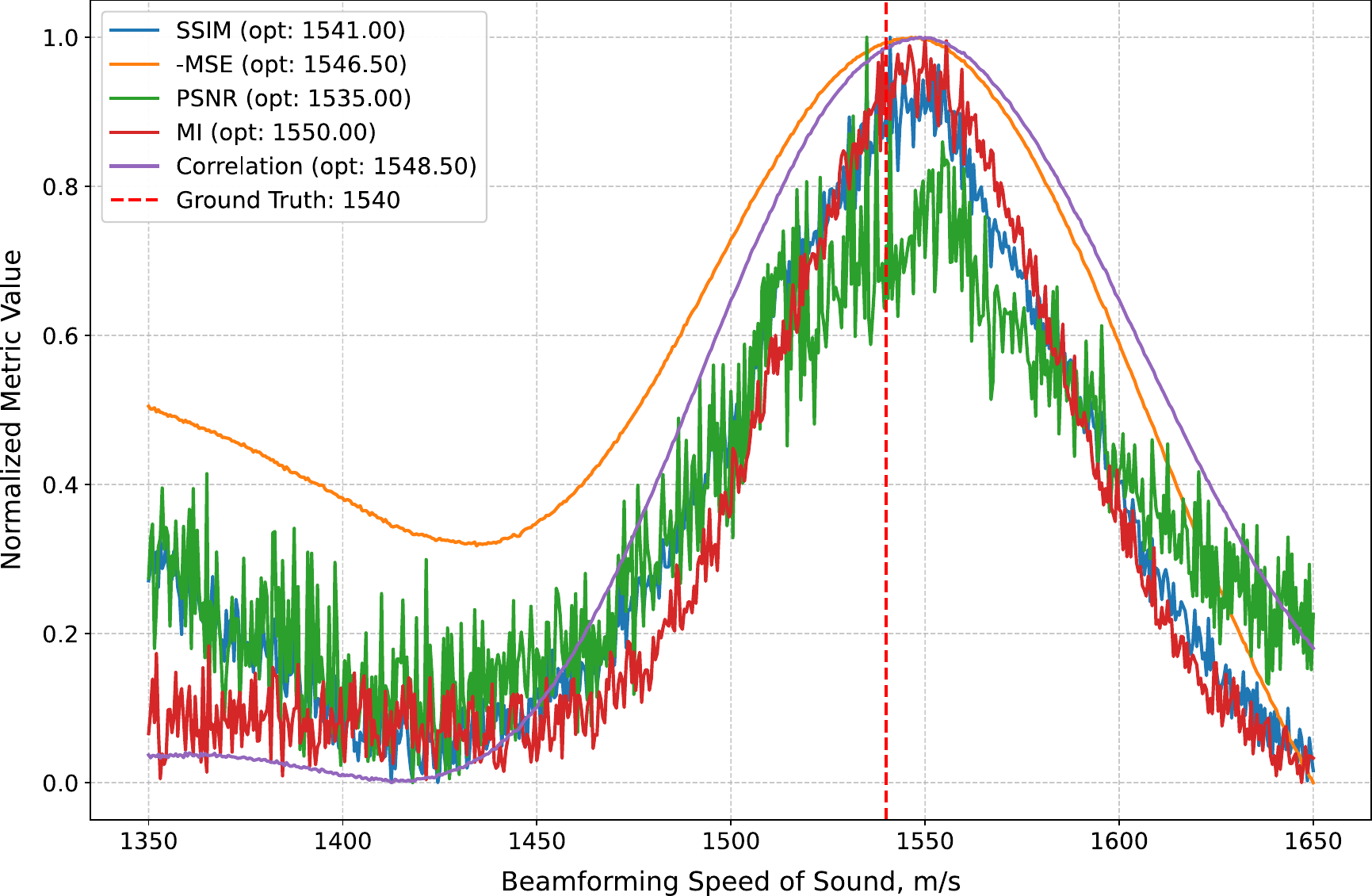}}
\caption{Sample behavior of image comparison metrics, demonstrated for Phantom 1, by normalizing each metric between their minimum and maximum for visualization purposes.
The optimum value (opt) found by each metric is indicated in the legend.}
\label{fig:metrics_behavior}
\end{figure}

For each Tx event, multitude of images were beamformed across the $n_\mathrm{s}$ test SoS values in $s_i\in[1450,1600]$\,m/s with a fine SoS increment of 0.5\,m/s to ensure a local optimum is not missed.
For this set of SoS values, having evaluated a metric $m(\cdot)$ using a set of one or more images $\{I_{n_\mathrm{t}}(s_i)\}$ depending on Single, Dual, or Full settings, an optimal global SoS value is then estimated by the given metric as $s^\star = \arg\max_{s_i} m(\{I_{n_\mathrm{t}}(s_i)\})$.
Sample patterns for the image comparison metrics are depicted in a normalized form in~\Cref{fig:metrics_behavior}.

\section{Results}
\subsection{SoS Estimation Accuracy}
We calculated SoS estimation errors by comparing any prediction with the known ground-truth of the corresponding experiment, \ie the reported absolute SoS error is $|s^\star-\hat{s}|$. 
For simulations, mean absolute error (MAE) and standard deviation of the three homogeneous phantoms are reported. 
For phantoms, the same statistics over the six data acquisitions per phantom are reported.
The results for each metric are seen comparatively in \Cref{tab:combinedData32}.

\begin{table*}
\centering
\caption{Absolute errors (mean$\pm$standard deviation) of global SoS estimation are reported for three Simulation realizations and and six acquisitions per Phantom in three settings using Single, Dual, Full (17) frames as input to the methods, where applicable. 
Error values higher than 75\,m/s (25\% of SoS range tested) are highlighted in red. 
For each experimental setting (column), the lowest error (and similarly the standard deviation) in each group is highlighted in bold, with the lowest across all the groups also being underlined. 
Processing time required to calculate a metric at a single beamforming SoS is listed in the last column.}
\label{tab:combinedData32}
\resizebox{\textwidth}{!}{
\begin{tabular}{|@{\,}c@{\,}|l|ccc|ccc|ccc|c|}
\cline{2-12}
\multicolumn{1}{c|}{} & \multirow{2}{*}{Method} & \multicolumn{3}{c|}{Simulations} & \multicolumn{3}{c|}{Phantom 1} & \multicolumn{3}{c|}{Phantom 2} & \multicolumn{1}{c|}{Time} \\ 
\multicolumn{1}{c|}{} &                  & Single       & Dual         & Full        & Single       & Dual         & Full         & Single       & Dual         & Full        & [ms]   \\ 
\hline
\multirow{5}{*}{\rotatebox{90}{Quality}} & Focus                  & \textcolor{red}{75.3 \tightpm \unc{26.2}}      & \textbf{21.0} \tightpm \textbf{\unc{11.2}}         & \textbf{7.0} \tightpm \unc{8.8}        & \textcolor{red}{178.8 \tightpm \unc{16.8}}       & \textbf{6.8} \tightpm \textbf{\unc{6.9}}       & \textbf{7.1} \tightpm \unc{6.3}        & \textcolor{red}{138.5 \tightpm \unc{2.1}}       & 17.8 \tightpm \textbf{\unc{4.7}}       & 11.3 \tightpm \unc{3.7}       & 16 \\
& Entropy                & \textcolor{red}{109.0 \tightpm \unc{81.6}}       & \textcolor{red}{132.7 \tightpm \unc{89.9}}       & 24.7 \tightpm \unc{12.9}        & \textcolor{red}{75.8 \tightpm \unc{46.4}}       & 44.2 \tightpm \unc{35.1}       & \textcolor{red}{99.3 \tightpm \unc{48.3}}       & \textcolor{red}{106.3 \tightpm \unc{30.1}}       & 70.9 \tightpm \unc{28.5}         & \textcolor{red}{99.0 \tightpm \unc{18.6}}        & 47 \\
& Tenengrad              & \underline{\textbf{11.0}} \tightpm \unc{3.9}        & 25.5 \tightpm \unc{13.4}        & 26.2 \tightpm \unc{9.5}        & \textcolor{red}{168.9 \tightpm \unc{13.8}}       & \textcolor{red}{186.6 \tightpm \unc{2.9}}       & \textcolor{red}{183.3 \tightpm \unc{5.8}}       & \textcolor{red}{153.5 \tightpm \unc{3.3}}       & \textcolor{red}{156.7 \tightpm \unc{1.5}}       & \textcolor{red}{154.0 \tightpm \unc{7.3}}      & 0.76 \\
& ANACVF              & \textcolor{red}{134.0 \tightpm \unc{98.8}}          & \textcolor{red}{136.8 \tightpm \unc{90.3}}         & 48.6 \tightpm \unc{6.3}        & \underline{\textbf{68.8}} \tightpm \underline{\textbf{\unc{42.5}}}       & \textcolor{red}{106.5 \tightpm \unc{4.0}}       & \textcolor{red}{107.5 \tightpm \unc{4.3}}       & \textcolor{red}{104.8 \tightpm \unc{14.7}}       & \textcolor{red}{87.8 \tightpm \unc{5.9}}       & \textcolor{red}{100.3 \tightpm \unc{13.1}}       & 5.7 \\
& ST-Ten              & 11.5 \tightpm \underline{\textbf{\unc{3.5}}}          & 23.2 \tightpm \unc{16.0}         & 7.7 \tightpm \textbf{\unc{4.3}}        & 69.8 \tightpm \unc{70.5}       & 17.0 \tightpm \unc{15.3}        & 8.0 \tightpm \textbf{\unc{3.4}}          & \textcolor{red}{85.6 \tightpm \unc{68.7}}       & \textbf{10.3} \tightpm \unc{9.6}         & \textbf{5.5} \tightpm \textbf{\unc{3.2}}        & 7.2 \\
\hline
\multirow{5}{*}{\rotatebox{90}{Comparison}} & SSIM                   & -            & \underline{\textbf{1.5}} \tightpm \unc{1.8}         & -           & -            & 9.3 \tightpm \unc{8.1}         & -            & -            & 6.1 \tightpm \unc{3.0}         & -           & 35 \\
& MSE                    & -            & 9.3 \tightpm \unc{2.5}         & -           & -            & 6.3 \tightpm \underline{\textbf{\unc{0.7}}}          & -            & -            & \underline{\textbf{4.6}} \tightpm \underline{\textbf{\unc{1.3}}}         & -           & 0.34 \\
& PSNR                   & -            & 19.2 \tightpm \unc{19.0}         & -           & -            & 11.0 \tightpm \unc{7.5}         & -            & -            & 7.7 \tightpm \unc{6.6}         & -           & 0.45 \\
& MI                     & -            & 4.5 \tightpm \unc{4.4}         & -           & -            & \underline{\textbf{5.8}} \tightpm \unc{2.5}         & -            & -            & 7.2 \tightpm \unc{3.0}          & -           & 37 \\
& Correlation            & -            & 2.7 \tightpm \underline{\textbf{\unc{0.6}}}       & -           & -            & 8.9 \tightpm \unc{1.4}         & -            & -            & 7.3 \tightpm \underline{\textbf{\unc{1.3}}}          & -           & 6.6 \\
\hline
& CV  & - & - & \underline{4.8} \tightpm \underline{\unc{1.2}} & - & - & \underline{4.3} \tightpm \underline{\unc{0.6}} & - & - & \underline{5.4} \tightpm \underline{\unc{0.7}}       & 62 \\
\hline
\end{tabular}}
\end{table*}

Since the reported errors are bounded by the SoS search range, we highlight in red the average errors that are larger than 25\% of the tested range, which likely include estimates at the bounds (\eg due to non-convex metric behavior) and are thus uninformative. 
The image comparison and multi-frame statistic methods are seen to perform similarly between Simulations and Phantoms, whereas the image-quality metrics demonstrate somewhat varying behaviour between these data domains.
In particular, the Tenengrad metric viable in Simulations does not generalize to Phantom data as well as the Full compounding for some metrics.
Overall, the assessed image quality metrics are not successful when a single image is used.
Nevertheless, when images compounded from multiple frames are used to increase SNR, Focus and ST-Ten become viable options for SoS estimation.
Compared to quality metrics, comparison metrics show a markedly superior performance in all experiments overall. 
Notably, MI and MSE emerge as effective options for global SoS estimation in phantom data, with approximately 5-7\,m/s accuracy. 
Regarding precision (measured via standard deviation of estimations), MSE and Correlation metrics are seen to perform well. 
Overall all given image-comparison metrics seem feasible for SoS estimation to different degrees.
The multi-frame statistic metric CV outperforms single-image quality metrics and performs on par with dual-frame image comparison metrics.
For homogeneous media, our best-performing image-comparison metrics attain absolute errors on the order of 5–7\,m/s using only two frames, while Focus and ST-Ten approach similar ranges only by Full compounding of many frames. 
Notably, image comparison metrics achieve their accuracy using only two frames, which can be acquired within microseconds to milliseconds depending on the transmit sequence.

In Table~\ref{tab:combinedData32} we also report average metric computation times for each metric for a single beamforming SoS value, calculated in Python using the NumPy package on an Intel Core i7-12700K CPU with 64GB RAM.
MSE is seen to be the fastest among all metrics, which makes it potentially the preferred metric of choice for global SoS estimation, given also its high accuracy, high precision, and relatively smooth metric behavior as observed in~\Cref{fig:metrics_behavior} for static scenes with large region of interest.

\subsection{Sensitivity to Estimation Window Size}
\begin{figure*}
\centerline{\includegraphics[width=1.0\textwidth]{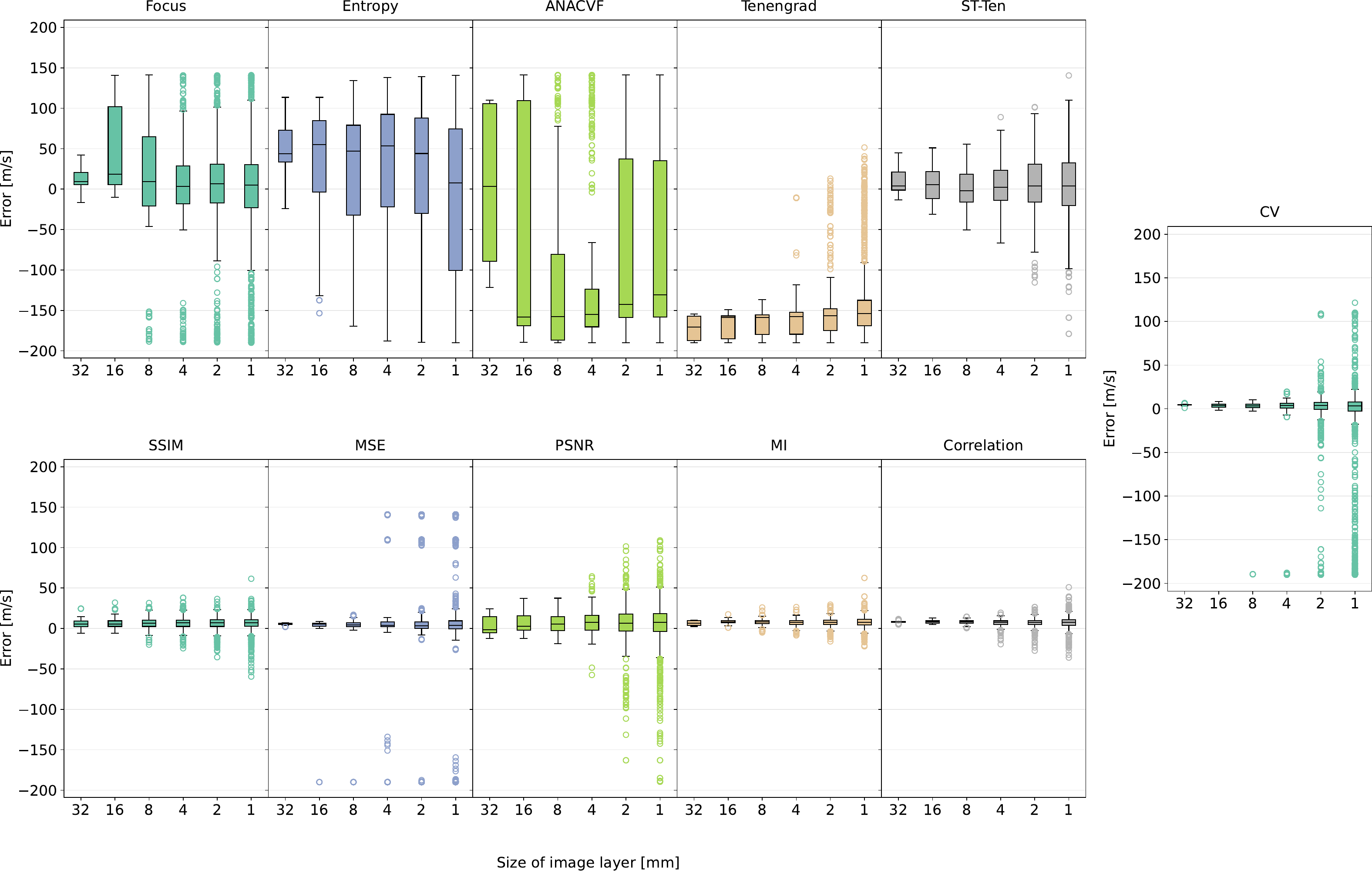}}
\caption{To study the sensitivity of image-based metrics to utilized image window, the distribution of SoS estimation errors are presented for using image patches (layers) of varying sizes from 32 down to 1\,mm in depth. 
Each layer location is perturbed vertically in increments of 0.1\,mm creating four layer positions each, in order to increase statistical stability of the evaluation. 
The evaluation is conducted for the two Phantoms.
For image comparison, the Dual acquisition setting was used via compounding for comparability with image similarity metrics.
Accordingly, the distributions for 32 summarize 48 estimations (2 phantoms $\times$ 6 acquisitions $\times$ 1 32\,mm-patch $\times$ 4 perturbations), whereas for 1 they summarize 1536 estimations containing 32 1\,mm-patches instead.}
\label{fig:slicing_full}
\end{figure*}

To further explore image metrics for estimations based on small regions of interest, we assessed them in image patches.
To that end, we subdivided the original 32\,mm image depth iteratively into smaller patches, \ie $\{2, 4, 8, 16, 32\}$ equi-depth layers of each $\{16, 8, 4, 2, 1\}$\,mm, respectively.
Then, using the image in each such patch, the optimal SoS and the corresponding error is found as described earlier. 
For a more robust evaluation, we employed multiple (herein 5) estimations per patch, including the original 32\,mm full image, by jittering its vertical location (layer depth) in increments of 0.1\,mm. 
This analysis aims to assess the stability and consistency of the metrics calculations. 
We conducted this evaluation on the Phantom data separately on the twelve acquisitions (six per phantom).
For the image quality metrics, Full compounding results are reported given their superior performance from the earlier experiment.
We report the results in \Cref{fig:slicing_full}, separated for each metric and patch (layer) size, with data points indicating the error given a patch and its vertical jittered location for each of the twelve acquisitions.

Focus and ST-Ten are again observed as the only potentially viable image quality options, while the image comparison metrics largely outperform the image quality metrics. 
Among comparison metrics, Mutual Information and Correlation perform consistently and with overall high accuracy for all tested layer sizes down to 1\,mm in our experimental setup. 
The performance of the multi-frame statistic metric CV is seen to deteriorate when the layer sizes are below 8\,mm.

\subsection{Evaluation in Layered Heterogeneous Medium}
\begin{figure*}
\centerline{\includegraphics[width=1.0\textwidth]{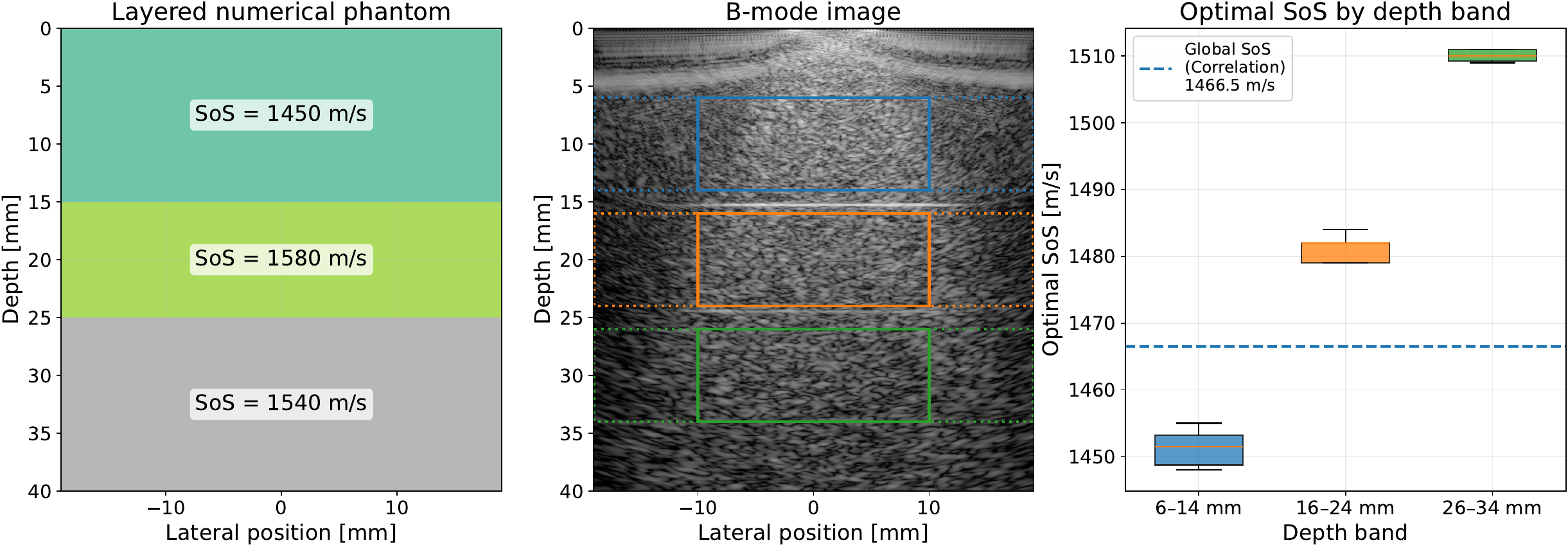}}
\caption{
Image-based SoS estimation within regions of interest (ROI) in a layered heterogeneous medium, demonstrated for the Correlation metric. 
Left: SoS map of the numerical phantom with three horizontal layers of 1450, 1580, and 1540~m/s. 
Middle: Corresponding B-mode image showing estimation ROIs within three depth bands (6-14~mm, 16-24~mm, and 26-34~mm). 
Within each band, SoS was estimated in a ROI of 20x8~mm$^2$, jittered in 1\,mm steps up to $\pm 9$\,mm laterally to assess robustness. 
Right: Distribution of SoS values estimated by maximizing Correlation within each ROI.
Due to cumulative time-delay effects, deeper bands are not expected to match the corresponding layer ground-truth; nevertheless, values show clear separation between the bands and low within-band variation indicates a high consistency. 
The dashed line indicates the global SoS with Correlation optimized over the entire image.}
\label{fig:layers_sos}
\end{figure*}

To assess metric behavior in heterogeneous tissue, we conducted an additional simulation with three horizontal layers of differing SoS values (1450, 1580, and 1540~m/s), as shown in~\Cref{fig:layers_sos}. 
Using the Correlation metric, we evaluated windowed SoS estimates within three depth bands (6-14~mm, 16-24~mm, and 26-34~mm), each sampled with multiple laterally-shifted subwindows to assess consistency. 
The resulting estimates show tight within-band distributions, indicating high measurement consistency at each depth. 
However, only the topmost layer yields estimates close to its true SoS, while deeper layers show systematic deviations due to the cumulative effect of overlying tissue on beamforming delays. 
These depth-dependent optimal SoS values can serve as input to layered SoS reconstruction methods such as \cite{bezek2025windowed}.

\subsection{Sensitivity to Motion}
Motion sensitivity has been evaluated on a tissue-mimicking phantom using a motorized probe holder with linear speeds of up to 1\,mm/s in lateral and axial directions, where the former induces image translation and the latter axial compression. 
Frame-rate of consecutive transmissions was limited by the data transport throughput, with each frame taking 37.5\,ms. 
Among the image quality metrics, Focus and ST-Ten were tested in \emph{Full} mode, based on their superior performance in stationary experiments (\Cref{tab:combinedData32}).
Therefore, the minimum range of motion is 0.0375\,mm for image similarity metrics requiring 2 transmissions, and 0.6375\,mm for image quality and multi-frame metrics, for which we used an acquisition loop of 17 transmissions.

\begin{figure*}
\centerline{\includegraphics[width=1.0\textwidth]{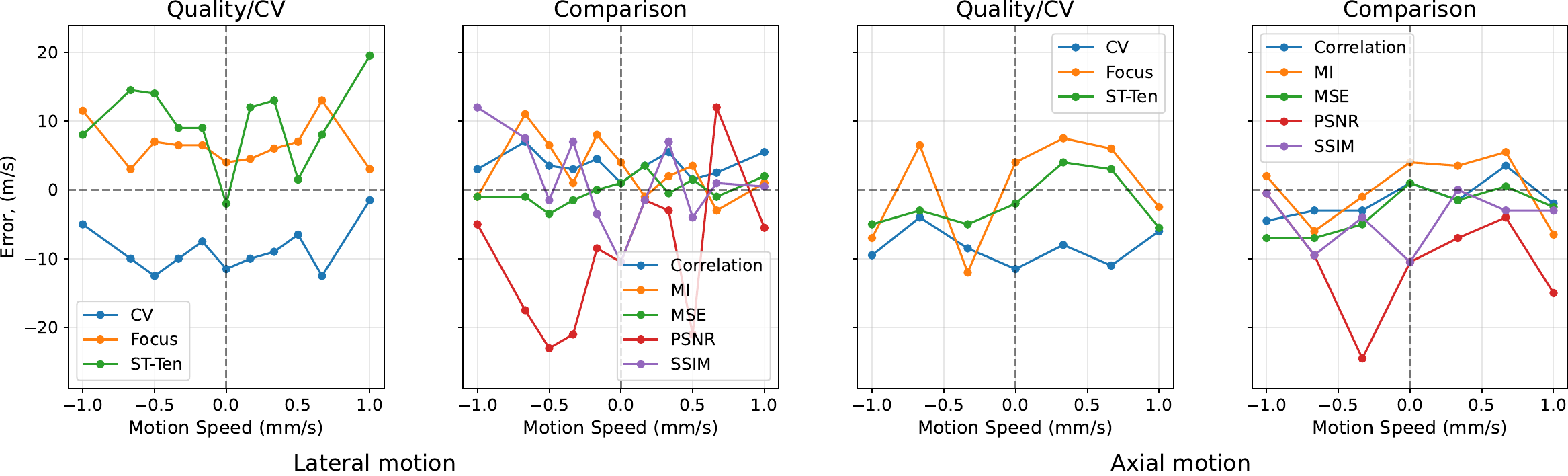}}
\caption{Motion sensitivity of global SoS estimation was evaluated by translating the ultrasound probe laterally and axially with speeds of up to 1\,mm/s using a motorized probe holder~\cite{schweizer2023robust}.}
\label{fig:motion}
\end{figure*}

Metrics performance in estimating known background SoS is depicted in \Cref{fig:motion}.
Absolute SoS errors remained modest across metrics. 
PSNR exhibited occasional single-measurement outliers up to 24.5\,m/s; while all other metrics stayed within a 12\,m/s absolute error envelope.
MSE and Correlation had overall lowest errors, both remaining under 7\,m/s throughout the motion tests.

Within this range of motion, none of the evaluated metrics displayed a systematically motion sensitive behaviour such as monotonic increase with speed either in signed or absolute value. 
Pairwise comparison metrics (except PSNR) were relatively robust to motion, despite much larger inter-frame motions imposed on them by the sequence constraints. 
Although individual metrics show variability, these are mostly within their typical random variations reported in \Cref{fig:slicing_full}, motion-originated errors in the recovered global SoS seem to be minimal for the tested settings.

\subsection{Global Optimality Assessment}
We further study the cases when a metric may have a local optimum at the sought GT SoS value, but with a global optimum somewhere far from this value, \eg at the bounds of the assessed SoS range.
This is relevant for the cases where the original evaluation shows a large error value, which could be reduced by considering a smaller SoS test range.
For this, we performed an additional evaluation within a tighter SoS range of $s_i\in\{\hat{s}\pm50\}$\,m/s around each known ground-truth SoS value. 
These additional results are tabulated in \Cref{tab:range50Data} in the Appendix and they are summarized visually in \Cref{fig:slicing_range50}. 
As seen, the metrics that have not performed satisfactorily in the earlier evaluation with a broad search range (\eg, MSE and PSNR for small layers, and Entropy and Tenengrad for all settings) do not perform well a smaller SoS search range either. 
The only appreciable improvement potentially is that the Single frame estimation becomes possible with under 20\,m/s accuracy using ST-Ten metric, despite with a relatively low precision.

\begin{figure*}
\centerline{\includegraphics[width=1.0\textwidth]{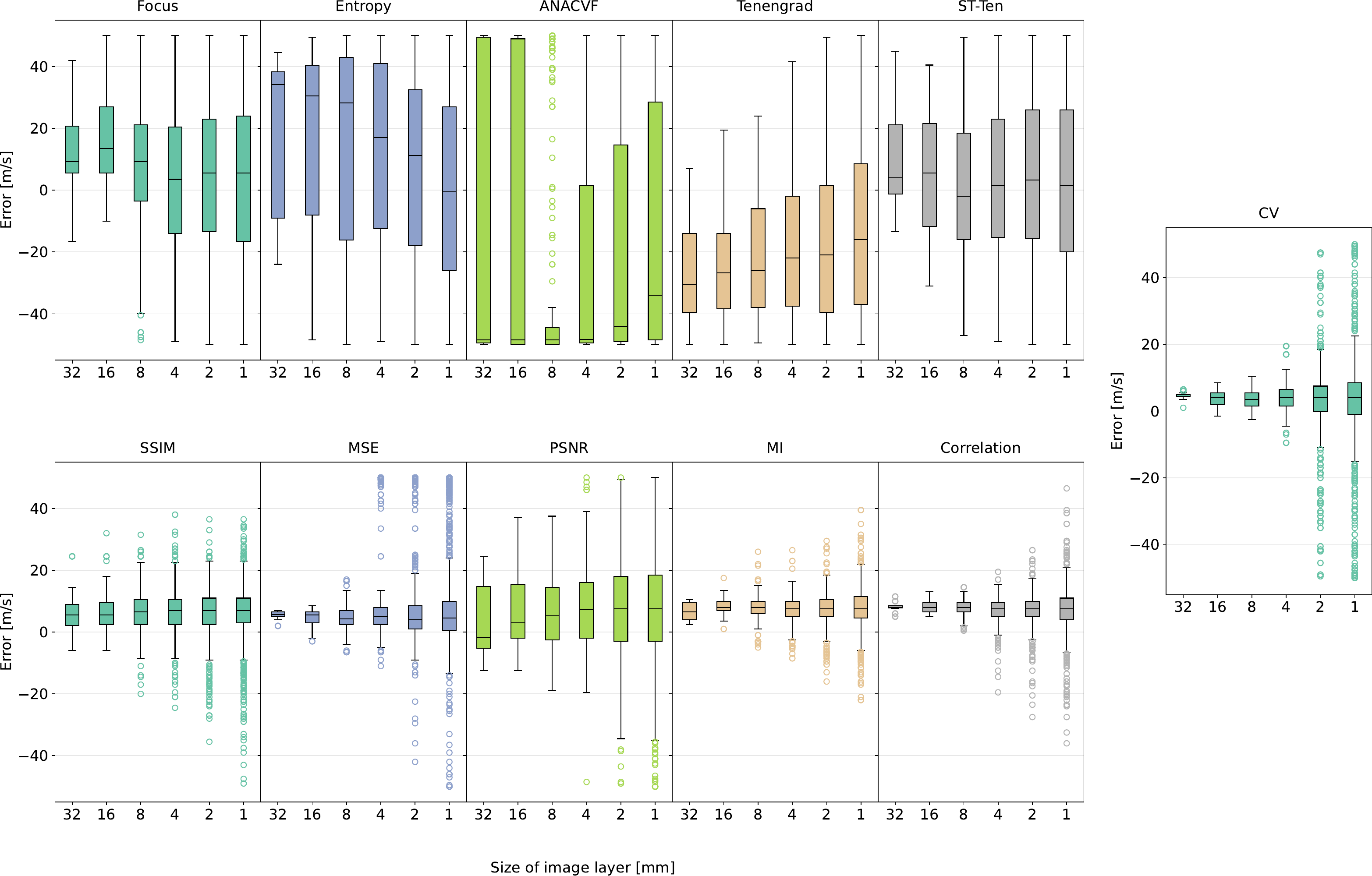}}
\caption{
Distributions of SoS estimation errors are presented for image patches (layers) of varying sizes from 32 down to 1\,mm in depth, similarly to \Cref{fig:slicing_full} but when the SoS search range is restricted around the known SoS for each experiment, \ie $s_i\in \{s_\mathrm{GT}\pm50\}$\,m/s. 
Each layer location is perturbed vertically in increments of 0.1\,mm creating four layer positions each, to present statistical variability. 
The evaluation is conducted for the two Phantoms.
For image quality, the Dual acquisition setting was used via compounding for comparability with image similarity metrics.
}
\label{fig:slicing_range50}
\end{figure*}

\subsection{Coarse SoS Search and Interpolation}
\label{sec:coarseSoS}
For comprehensive and robust evaluation by capturing global optima, all methods were evaluated with a small beamforming SoS step of 0.5\,m/s, requiring many costly beamforming operations for each Tx event.
To test the metrics on coarser SoS grids, we used 20\,m/s steps (\ie, 12 images per Tx to beamform) by quadratic interpolation for sub-sampling around the maximum to refine optimal SoS values.
The resulting SoS estimations were within standard deviation range of the values reported in~\Cref{tab:combinedData32}, showing that a coarser SoS trial-and-error grid does not cause substantial performance reduction.

\subsection{In Vivo study for a clinical task}
To test the feasibility of the presented metrics given the heterogeneity of tissues, lower SNR of clinical acquisitions, and potential motions of operator and patients, we next conducted an in vivo clinical study.
Note that 
Note that the SoS value optimized by the metrics is an optimal beamforming SoS over the chosen region of interest and any image locations superficial to that region where the acoustic waves travel through~\cite{bezek2025windowed}, which need not equal a physical mean SoS.
Nevertheless, for relatively homogeneous tissue regions, such global SoS would approximate the average tissue SoS.

We evaluated the presented image-based metrics on a dataset collected for breast density (BD) classification.
High BD is a potential cancer risk~\cite{BODEWES202262} and also reduced sensitivity of mammography to detect cancer~\cite{kolb2002comparison}, so the knowledge of BD can allow clinical stratification.
Also, reporting to patients their BD has recently been imposed by FDA~\cite{bezek2025breast}.
Hence estimation of BD via ultrasound bears high clinical relevance.
For evaluation, data from 92 patients who underwent mammography imaging 
as well as ultrasound breast examination during standard clinical diagnostic procedures were used. 

Since the ground-truth values for tissue SoS are not known, we assessed the metrics via comparisons to a silver-standard by a state-of-the-art model-based SoS estimation method (g-SoS) presented in \cite{bezek2025breast}.
Each metric was compared to the 92 g-SoS values using mean signed error (ME) and mean absolute error (MAE), as well as rank (Spearman) and linear (Pearson) correlations.
\Cref{tab:reference_metric_comparison} tabulates these results with the metrics in descending order of their rank correlation, \ie patient SoS comparative order being most similar to g-SoS.

\begin{table*}
\centering
\caption{Comparison of optimal beamforming SoS to reference g-SoS, using Spearman and Pearson correlations ([-1,1]), as well as mean signed error (ME) and mean absolute error (MAE) with their standard deviations.
Metrics are sorted in descending order of their rank correlation, with correlations above 0.5 shown in bold.}
\label{tab:reference_metric_comparison}
\resizebox{.7\textwidth}{!}{
\begin{tabular}{|l|rrrr|}
\hline
Metric & Spearman $\rho$ & Pearson r & ME [m/s] & MAE [m/s] \\
\hline
\textbf{Correlation}   & \textbf{0.872} & \textbf{0.820} & \textbf{-6.7\tightpm22.2}   & \textbf{16.4\tightpm16.4}\\
\textbf{ST-Ten/Full}   & \textbf{0.862} & \textbf{0.861} & \textbf{-22.7\tightpm17.4}  & \textbf{24.0\tightpm15.6}\\
\textbf{Focus/Full}    & \textbf{0.842} & \textbf{0.818} & \textbf{0.0\tightpm26.2}   & \textbf{19.5\tightpm17.3}\\
PSNR          & 0.401 & 0.361 & -12.4\tightpm49.7  & 37.1\tightpm35.2\\
MI            & 0.394 & 0.284 & 11.4\tightpm67.8   & 46.4\tightpm50.5\\
SSIM          & 0.389 & 0.372 & -9.0\tightpm57.3   & 42.9\tightpm38.8\\
ST-Ten/Dual   & 0.274 & 0.339 & -50.4\tightpm39.3  & 50.5\tightpm39.1\\
Focus/Dual    & 0.134 & 0.162 & 48.0\tightpm91.7   & 87.4\tightpm54.9\\
Focus/Single  & -0.007 & -0.129 & 87.6\tightpm105.2  & 131.9\tightpm35.5\\
CV            & -0.144 & 0.061 & -76.9\tightpm46.9  & 77.6\tightpm45.7\\
MSE           & -0.148 & -0.131 & -52.8\tightpm56.2 & 55.7\tightpm53.3\\
ST-Ten/Single & -0.161 & -0.109 & -86.7\tightpm40.9 & 86.7\tightpm40.9\\
\hline
\end{tabular}}
\end{table*}

\begin{figure*}
\centerline{\includegraphics[width=1.0\textwidth]{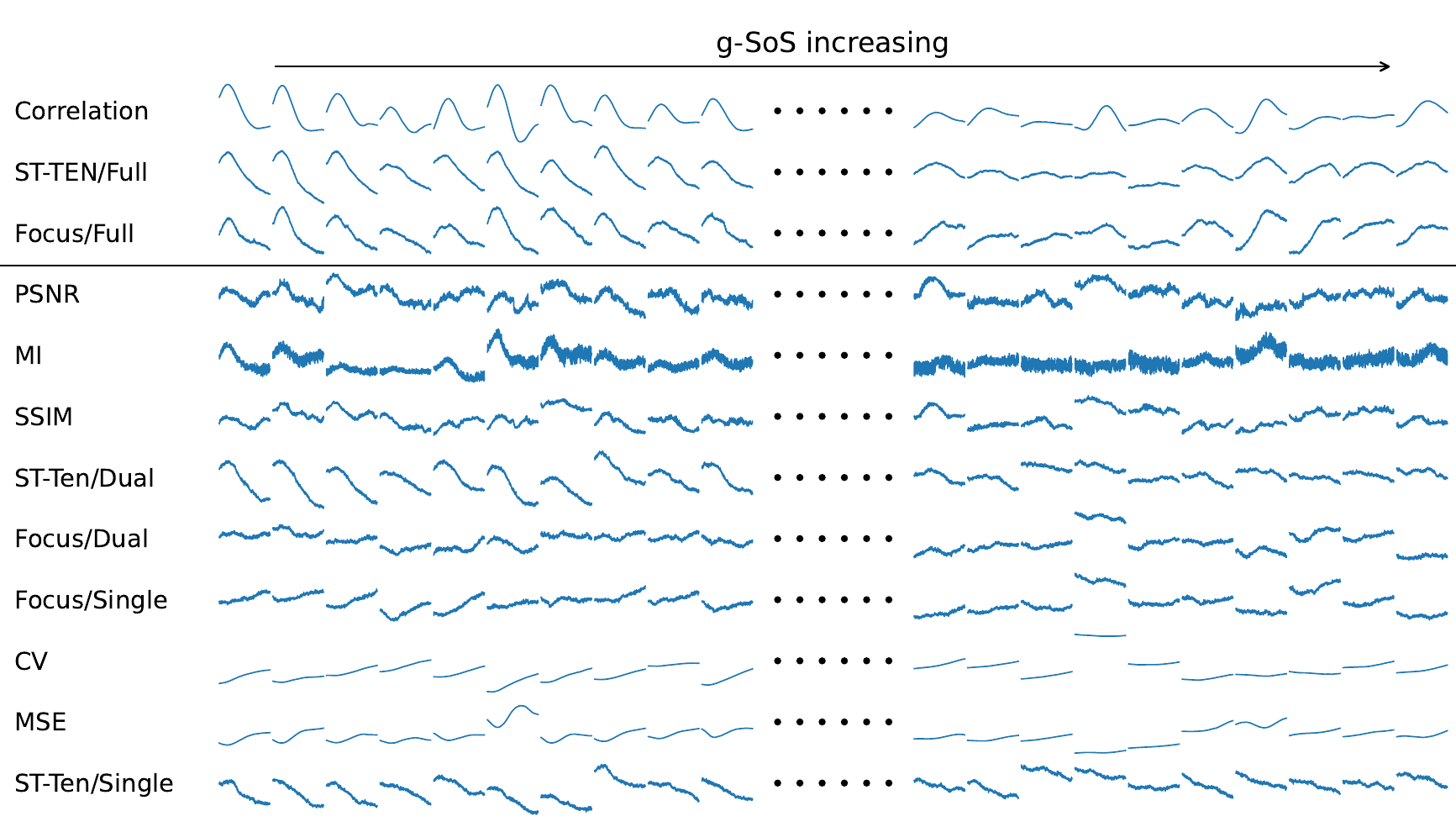}}
\caption{Metric profiles across the tested SoS range for the patients with the lowest and highest reference SoS values (g-SoS). 
Each row corresponds to a metric (\Cref{tab:reference_metric_comparison}) with the line separating less reliable ones, \ie with correlations to reference below 0.5. 
Curves show metric values while sweeping the beamforming SoS in 0.5\,m/s steps, similarly to \Cref{fig:metrics_behavior}.
Patients are sorted in ascending g-SoS, with the profiles displayed for the lowest and the highest 10 cases. 
}
\label{fig:clinical_metric_behavior}
\end{figure*}

\begin{figure*}
\centerline{\includegraphics[width=1.0\textwidth]{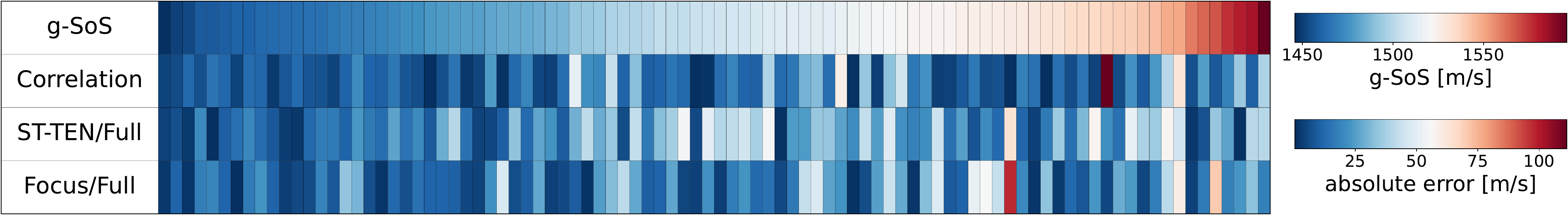}}
\caption{Absolute errors $|s^*-s_{\mathrm{g\text{-}SoS}}|$ between our image-based estimated SoS $s^*$ and the reference global SoS $s_{\mathrm{g\text{-}SoS}}$ for the three best-performing metrics (Correlation, ST-TEN/Full, and Focus/Full). Columns correspond to patients sorted by increasing g-SoS, shown in the top row.  
}
\label{fig:error_behavior}
\end{figure*}

Only Correlation (Dual), ST-Ten, and Focus (Full) achieved meaningful correlation values of~$>$$0.5$, with MAE and SD $<$$25$\,m/s.
These corroborate earlier observations.
All other metrics perform significantly worse in errors, correlation, or both.
To better understand this performance discrepancy, we examined the metric behavior across the SoS search range for individual patients, as shown in \Cref{fig:clinical_metric_behavior}. 
For well-performing metrics (Correlation, ST-Ten/Full, Focus/Full), the metric curves typically exhibited a clear dominant maximum near the reference g-SoS, although the peaks somewhat flatten at increasing g-SoS with the optima less distinct and potentially further from g-SoS.
Lower-performing metrics, however, frequently yielded nonconvex or monotonically increasing/decreasing profiles, with optima occurring at the boundaries of the tested SoS range rather than at physiologically meaningful values. 
This was true also for some metrics such as MSE and CV that showed satisfactory performance in the synthetic and phantom experiments, suggesting that metrics may have differing levels of robustness to large tissue heterogeneity and lower SNR characteristics of in vivo data.
Furthermore, as seen in \Cref{fig:error_behavior} for the better performing three metrics, the estimation accuracy overall decreased with increasing g-SoS, \ie denser breasts that are more heterogeneous.

\begin{figure*}
\centerline{\includegraphics[width=0.8\textwidth]{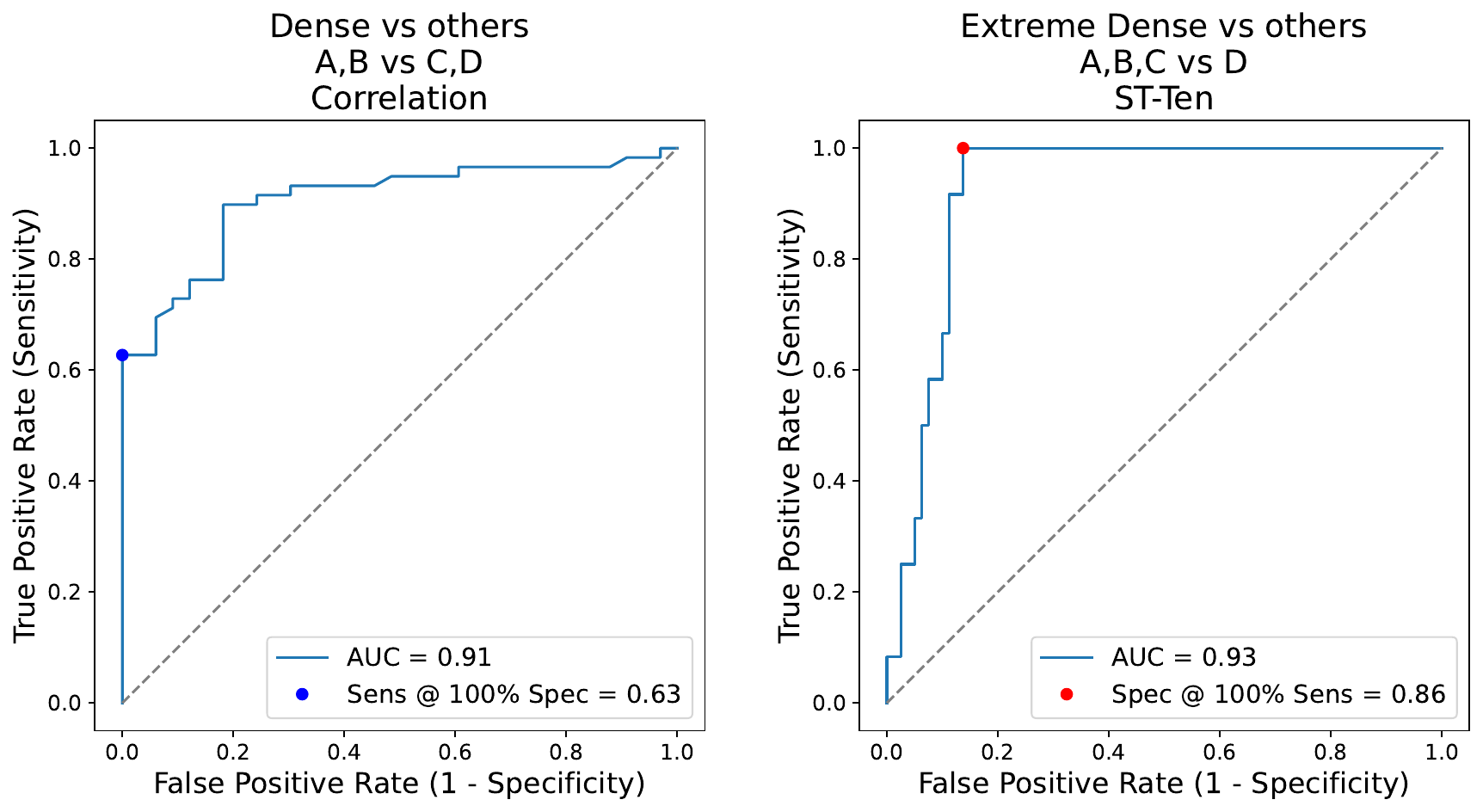}}
\caption{Receiver operating characteristic (ROC) curve for Correlation in classifying dense (m-ACR classes C and D) breasts and ST-Ten for extremely dense (m-ACR class D) breasts.}
\label{fig:ROC_curve}
\end{figure*}

Next we used the SoS predictions by the best performing image-based metric (Correlation) for classifying breast density based on BI-RADS categories (A/B/C/D) from gold-standard annotations of the 92 patient mammograms by expert radiologist as described in~\cite{bezek2025breast}.
We tested the classification of dense (m-ACR classes C and D) from non-dense (A and B) breasts, as well as extremely dense (D) from all others (A,B,C).
Receiver operating characteristic (ROC) curves are shown in \Cref{fig:ROC_curve}, which quantifies true and false positive rates by sweeping SoS thresholds for a classification decision.
The results are reported as the area under curve (AUC) of these ROCs, and the ability for most cases that can be successfully excluded (\ie, highest specificity at 100\% sensitivity, \emph{spec@100\%sens}) or included (\ie, highest sensitivity at 100\% specificity, \emph{sens@100\%spec}).
For dense breast (C\&D vs. A\&B) binary classification, Correlation achieved an AUC of 0.91 with 63\% sens@100\%spec, while the reference g-SoS method was reported with an AUC of 0.931 and 78\% sens@100\%spec in~\cite{bezek2025breast}. 
Similarly for extremely dense breast (D) classification, ST-Ten (Full) has achieved AUC of 0.93 with 86\% spec@100\%sens, while the reference g-SoS method has an AUC of 0.901 and 77.5\% spec@100\%sens~\cite{bezek2025breast}.

\section{Discussion}

We have presented an extensive analysis of image-based metrics for model-free estimation of global SoS in simulations and phantoms.
The metrics are categorized into image quality metrics that check focusing quality by processing single (including compounded) B-mode images, image comparison metrics that measure the similarity/concordance between two frames, and multi-frame statistical metrics that evaluate the group-wise concordance via statistical techniques.

Summarizing our results while focusing on the phantom evaluations with homogeneous medium, none of the assessed image quality metrics performed satisfactorily when applied to a single image frame.
Compounding two or more frames improved performance, but still only the Focus and ST-Ten metrics attained a reasonable level of success.
Comparatively, image comparison and statistical metrics performed substantially better across all the analyzed experimental scenarios. 
Notably, SSIM, CV, MSE, MI, and Correlation emerged as effective metrics in analyzing large image patches in static frames from Simulations and Phantoms. 
For smaller patch (layer) sizes in static Phantom images, MI and Correlation yielded performance superior to all the other considered metrics, including CV and MSE.
When applied on an in vivo clinical task, only Correlation, ST-Ten and Focus (in Full aggregation mode) have performed well, achieving high correlation with a current state of the art model-based method g-SoS.
The benefit of compounding for image quality metrics can be attributed to averaging out path-dependent variations: since each Tx event produces sound waves traveling through different acoustic paths, compounding multiple frames reduces sensitivity to these path-specific effects while preserving the global defocusing artifacts caused by SoS mismatch. Furthermore, since incorrect SoS assumption already causes internal defocusing within each frame, compounding frames that are additionally misaligned with respect to each other amplifies this blurring, increasing the sensitivity of quality metrics to SoS errors. Together, these effects may explain why Focus/Full and ST-Ten/Full achieved relatively better performance in the in vivo evaluation, where tissue heterogeneity introduces substantial path-dependent variations.
When compared to gold-standard radiologist annotations on a breast-density classification task, some of these metrics, in particular Correlation and ST-Ten (Full), could even marginally outperform g-SoS.
Such in vivo results demonstrate potential clinical utility of global SoS estimation despite its homogeneous-equivalent approximation of inherently heterogeneous tissues.
Surprisingly, in vivo performance of MSE, MI, and CV were relatively poor in contrast to their promising results from some of the earlier experiments on Simulations and Phantoms, which might be explained by their higher sensitivity to heterogeneity and lower SNR of in vivo data.

\begin{figure*}
\centerline{\includegraphics[width=1.0\textwidth]{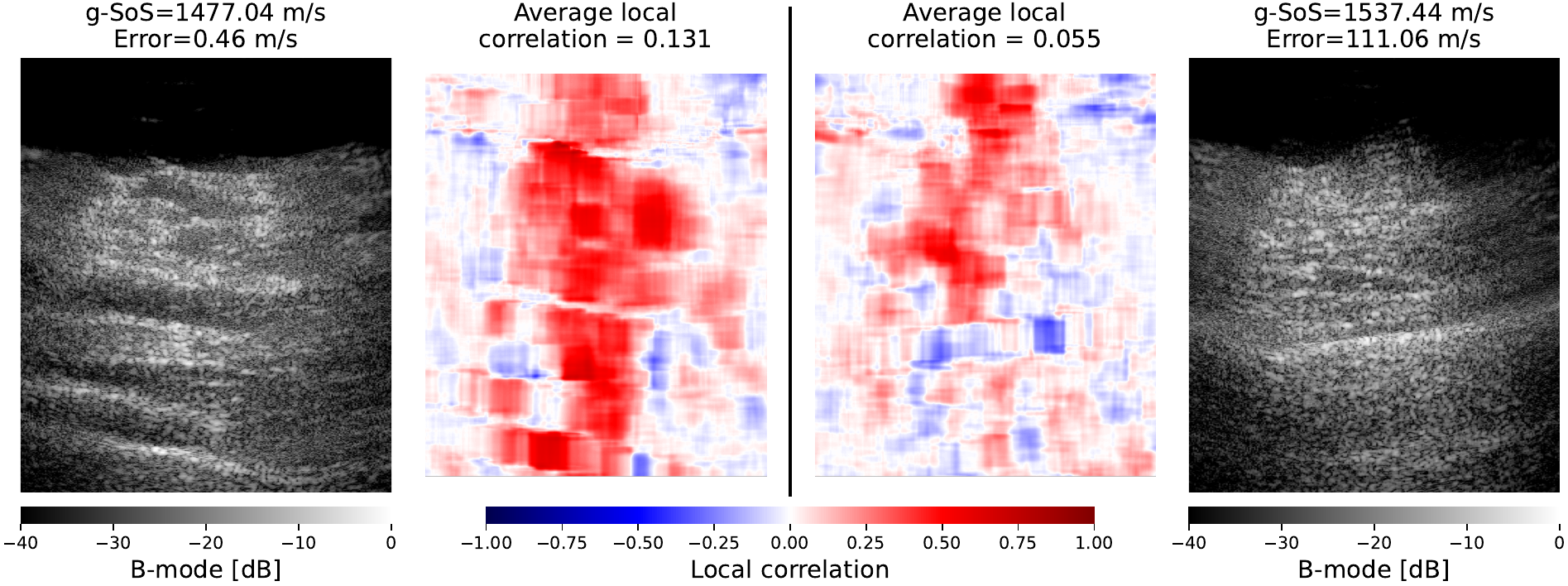}}
\caption{Spatial analysis of the Correlation metric for two two breast samples with the smallest (left) and largest (right) absolute errors with respect to the reference g-SoS. 
B-mode images are beamformed at the reference g-SoS values.
Correlation maps show the local correlation with a $3\times 3$\,mm sliding window (comparable to PSF size) between the frame pair both beamformed at the (full frame) Correlation metric estimated SoS value.
The reported \emph{average correlation} is the mean value across the correlation map. 
}
\label{fig:local_correlation}
\end{figure*}

To study the potential failure mechanisms with the in vivo data, we analyzed the best-performing metric (Correlation) for the extreme cases of minimum and maximum error with respect to the reference g-SoS.
To relate the metric variation spatially to tissue observed, we computed local (windowed) correlations as shown in~\Cref{fig:local_correlation}.
The image with a low estimation error exhibits a larger region with high local correlations, indicating prominent tissue structures that can be reliably matched between the transmit events. 
Conversely, the image with a high error shows most image regions having minimal correlation, particularly at increasing depths, consistent with weaker and less reliable optima in metric profiles. 
We hypothesize that Correlation remains sensitive to structural content and detail, whereas pixel-wise metrics like MSE and CV become dominated by speckle noise and cumulative sound propagation artifacts that may not carry SoS-dependent information. 

Image comparison and multi-frame statistical metrics can take arbitrary signals as input (pixel intensities), therefore we used post-beamformed RF data as a high-resolution input for them.
Image-quality metrics, on the other hand, mostly rely on the visual appearance based indicators, such as image gradients, so they cannot function reliably on RF signals given their modulated nature; therefore we used (envelope-detected) B-mode images as their input.
Although these still contain characteristic US speckle noise, it is seen from the results that some image-quality metrics, including Focus and our US-adapted ST-Ten, can still perform well to some extent in certain settings such as using compounded frames.
Their performance is indeed satisfactory even for processing smaller image patches.
These metrics hence become relevant when access to low-level RF data is not possible, as they operate on B-mode images alone. 
Although a way to control the beamforming SoS (that the B-mode images are generated by) would still be necessary, such dynamic control is becoming more and more available in ultrasound systems, \eg through a user-exposed knob or programatically~\cite{bezek2025sound}, which can then be controlled for testing multiple SoS values on-the-fly as studied in this paper.

A major advantage of image-based metrics studied herein is their independence of any transmission sequence of choice. We studied diverging wave (virtual source) acquisitions in particular for compatibility of the final evaluations with the clinical data obtained earlier~\cite{schweizer2023robust}. Nevertheless, some of these metrics have been applied successfully also with focused transmission in ~\cite{bezek2025sound} on a point-of-care device.

If a fast computation is desired, MSE can be used within relatively large image patches, \ie regions of interest.
As the computation times are reported for full 32\,mm image window, processing smaller patches would take shorter processing times.
Moreover, although we performed our experiments with a very fine SoS sampling increment of 0.5\,m/s not to miss any local behaviour/maxima, one can in practice sample SoS much coarser and then apply some subsampling assuming metric smoothness.
Indeed, we conducted an additional experiment with a very coarse beamforming SoS step of 25\,m/s followed by quadratic curve-fitting near the maxima:
The results shown for Simulations and Phantoms in \Cref{tab:combinedData32_interp25} in the Appendix demonstrate that the accuracy of most metrics do not change significantly, therefore indicating the potential for performing metric optimizations with limited computational resources.
The overall time required for SoS estimation can be expressed as $T_{\text{total}} \approx n_t \cdot t_{\text{acq}} + n_s \cdot (n_t \cdot t_{\text{bf}} + t_{\text{metric}})$, where $n_t$ is the number of Tx events (1-17 for the metrics/experiments presented herein), $t_{\text{acq}}$ is the acquisition time per Tx event (on the order of $2 \times \text{depth}/\text{SoS}$, i.e., microseconds), $n_s$ is the number of SoS trials, $t_{\text{bf}}$ is the beamforming time per frame, and $t_{\text{metric}}$ is the metric computation time (\Cref{tab:combinedData32}). 
In our GPU-based implementation, beamforming required approximately 1.5~ms per frame. 
Although the total time was dominated by the SoS trial attempts $n_s$; however, we achieved comparable accuracy (\Cref{sec:coarseSoS}) using only 12 trials (25~m/s steps with interpolation), which substantially reduces the computational time (down to around 100 ms for Correlation in a nonoptimized implementation). 
Note that SoS computations can be run sporadically, e.g., every few seconds, and after data acquisition, its computations may also be performed in parallel to other (e.g., B-mode) imaging.

The hyper-parameters for ST-Ten (frequency bands and Gaussian kernel size) were intentionally optimized on simulation data, rather than on phantom or in vivo data, to avoid overfitting to specific hardware characteristics. Although ST-Ten with such generic parametrization is found to perform successfully in our experiments, deployment in practice could further benefit from finetuning these parameters, \eg on a phantom with known SoS using the intended hardware and acquisition setting.

As in many estimation problems, repeatability (\ie precision, represented herein by standard deviation) of measurements is arguably more important and relevant than absolute errors.
This is first due to the fact that the actual SoS of samples in the experimental conditions may not be exactly as they are originally declared, due to the temperature differences or the phantoms drying over time.
Furthermore, various assumptions in beamforming such as the Tx SoS setting and the fixed time offset (often referred as $t0$) may bias the identified SoS value.
Nevertheless, given highly repeatable measurements also sensitive to SoS variations, it is often possible to calibrate such absolute measurement errors, \eg using independently measured SoS.
Furthermore, for differential evaluations, \eg for diagnostic purposes, actual SoS values may be less relevant compared to their precise and repeatable differentiation capability (except for a standardization effort in the long run, \eg across multiple systems or setups).
Note that MI and Correlation both provide relatively high precision in Phantoms, even in small image patches.
Although CV demonstrates marginally higher precision for the entire frame, its repeatability rapidly reduces for smaller frames (even for half frame of size 16\,mm) as observed in \Cref{fig:slicing_full}, where the CV estimates (boxplot medians) are seen to vary across different image sizes.
In contrast, Correlation preserves the median across images sizes, indicating promise for extending to layered estimations as in~\cite{bezek2025windowed}.
Concerns of possible motion impact have been addressed in a separate experiment with linear motion in lateral and axial directions, which has shown no visible motion sensitivity of metrics. In vivo evaluation was also performed on the data acquired with a handheld ultrasound probe that inherently involves motion artifacts.

Taken together, these findings offer a complementary perspective to studies emphasizing best-case point estimates obtained with coarser grid spacing. 
By employing finer SoS sampling and repeated acquisitions, our design aims to characterize precision and stability of the recovered optima. 
Reported errors are therefore not directly interchangeable across protocols, but appear broadly consistent in magnitude once differences in sampling granularity and reporting practice are considered~\cite{napolitano2006sound,he2009sound,he2017sound,benjamin2018surgery,xiao2024realtime}.

\section{Conclusions}
We have studied different image-based metrics for estimating the global SoS in simulation and phantom data as well as evaluated its applicability to clinically-relevant in vivo settings. 
In particular, our two proposed metrics, ST-Ten and Focus, demonstrated remarkable performance among image quality metrics, both in global and layered estimation scenarios as well as for in vivo evaluation. 
Image comparison metrics demonstrated performance superior to image quality metrics in most of simulation and phantom scenarios evaluated, while Correlation specifically has maintained high performance when applied to in vivo data on par with ST-Ten and Focus.  
Multi-frame statistic metric Coefficient of Variation (CV) performed well for large image areas, but yielded inferior results when processing smaller image patches and in vivo data, potentially due to its sensitivity to low SNR.
This approach also requires multiple (many) frames, the acquisition and processing of which both incur longer times. 
Image-based metrics are shown to be an effective alternative to physical model-based approaches, minimizing reliance on model accuracy and robustness.
Future research could further explore these metrics in vivo, implementing them on general US systems such as point-of-care devices~\cite{bezek2025sound}, extending them for resolving individual layer SoS values~\cite{bezek2025windowed}, or exploiting their potential in improving ultrasound image quality and resolution.
Additionally, learned methods could potentially be trained to directly predict optimal SoS from image features, bypassing the trial-and-error search entirely; the metrics and datasets studied herein could serve as training targets and benchmarks for such approaches.

\section{Acknowledgments}
Funding was provided by the Centre for Interdisciplinary Mathematics and the Medtech Science and Innovation Centre at Uppsala University in Sweden.
The authors thank the staff at Kantonsspital Baden, particularly Rahel A. Kubik-Huch, Monika Farkas, Anna Potempa, Cornelia Leo, Silke Callies, for their support with the clinical study.
The authors thank Dieter Schweizer for his instrumental role in setting up the data acquisition setup and in the clinical study.
The authors thank Can Deniz Bezek for his invaluable support and advice throughout this research.

\section{Declaration of generative AI and AI-assisted technologies in the manuscript preparation process.}
During the preparation of this work, ChatGPT (Open AI) and Claude.ai (Anthropic) were used to improve grammar, find synonyms, and typographic checking. Afterwards, the authors reviewed and edited the text and take full responsibility for the content of the published article.

\bibliographystyle{elsarticle-num}
\bibliography{references}

\clearpage
\appendix
\section{Appendix}

\begin{table}[H] 
\centering
\caption{Local optimality study of the metrics by restricting the SoS search range around the known SoS for each experiment, \ie $s_i\in \{c_\mathrm{GT}\pm50\}$\,m/s.
Absolute errors (mean$\pm$standard deviation) of global SoS estimation are reported in three datasets for using Single, Dual, and Full (17) frames as input to the methods, where applicable. 
Error values higher than 25\,m/s (25\% of SoS range tested) are highlighted in red. 
For each experimental setting (column), the lowest error (and similarly the standard deviation) in each group is highlighted in bold, with the lowest across all the groups also being underlined.
}
\label{tab:range50Data}
\resizebox{\textwidth}{!}{
\begin{tabular}{|@{\,}c@{\,}|l|ccc|ccc|ccc|}
\cline{2-11}
\multicolumn{1}{c|}{} & \multirow{2}{*}{Method} & \multicolumn{3}{c|}{Simulations} & \multicolumn{3}{c|}{Phantom 1} & \multicolumn{3}{c|}{Phantom 2}  \\ 
\multicolumn{1}{c|}{} &                  & Single       & Dual         & Full        & Single       & Dual         & Full         & Single       & Dual         & Full  \\ 
\hline
\multirow{5}{*}{\rotatebox{90}{Quality}} & Focus                  & \textcolor{red}{50.0 \tightpm \unc{0.0}}      & \textbf{21.0} \tightpm \textbf{\unc{11.2}}         & \textbf{7.0} \tightpm \unc{8.8}        & \textcolor{red}{45.9 \tightpm \unc{2.4}}       & \textbf{6.8} \tightpm \textbf{\unc{6.9}}       & \textbf{7.1} \tightpm \unc{6.3}        & \textcolor{red}{39.6 \tightpm \unc{8.1}}       & 17.8 \tightpm \textbf{\unc{4.7}}       & 11.3 \tightpm \unc{3.7}       \\
& Entropy                & 16.2 \tightpm \unc{8.8}       & \textcolor{red}{40.7 \tightpm \unc{10.5}}       & 24.7 \tightpm \unc{12.9}        & \textcolor{red}{20.3 \tightpm \unc{14.8}}       & 21.4 \tightpm \unc{9.4}       & \textcolor{red}{33.7 \tightpm \unc{7.1}}       & \textcolor{red}{33.8 \tightpm \unc{12.6}}       & \textcolor{red}{38.9 \tightpm \unc{4.3}}         & \textcolor{red}{33.1 \tightpm \unc{15.0}}        \\
& Tenengrad              & \underline{\textbf{11.0}} \tightpm \unc{3.9}         & \textcolor{red}{25.5 \tightpm \unc{13.4}}         & \textcolor{red}{26.2 \tightpm \unc{9.5}}        & \textcolor{red}{41.3 \tightpm \unc{11.9}}       & \textcolor{red}{30.3 \tightpm \unc{15.1}}       & \textcolor{red}{49.4 \tightpm \unc{1.1}}       & \textcolor{red}{41.0 \tightpm \unc{6.1}}       & 25.0 \tightpm \unc{13.0}       & \textcolor{red}{47.8 \tightpm \unc{3.7}}      \\
& ANACVF              & \underline{\textbf{11.0}} \tightpm \underline{\textbf{\unc{2.0}}}          & 23.3 \tightpm \unc{18.4}         & \textcolor{red}{46.5 \tightpm \unc{3.9}}        & \textcolor{red}{34.8 \tightpm \unc{15.6}}       & \textcolor{red}{47.2 \tightpm \unc{5.2}}       & \textcolor{red}{49.0 \tightpm \unc{1.2}}       & \textcolor{red}{44.0 \tightpm \unc{6.1}}       & \textcolor{red}{49.3 \tightpm \unc{0.6}}       & \textcolor{red}{49.8 \tightpm \unc{0.2}}       \\
& ST-Ten              & 11.5 \tightpm \unc{3.5}          & 23.2 \tightpm \unc{16.0}          & 7.7 \tightpm \textbf{\unc{4.3}}       & \underline{\textbf{18.3}} \tightpm \underline{\textbf{\unc{14.1}}}       & 17.0 \tightpm \unc{15.3}        & 8.0 \tightpm \textbf{\unc{3.4}}          & \underline{\textbf{19.5}} \tightpm \underline{\textbf{\unc{18.8}}}       & \textbf{10.3} \tightpm \unc{9.6}         & \textbf{5.5} \tightpm \textbf{\unc{3.2}}        \\
\hline
\multirow{5}{*}{\rotatebox{90}{Comparison}} & SSIM                   & -            & \underline{\textbf{1.5}} \tightpm \unc{1.8}         & -           & -            & 9.3 \tightpm \unc{8.1}         & -            & -            & 6.1 \tightpm \unc{3.0}         & -           \\
& MSE                    & -            & 9.3 \tightpm \unc{2.5}         & -           & -            & 6.3 \tightpm \underline{\textbf{\unc{0.7}}}          & -            & -            & \underline{\textbf{4.6}} \tightpm \underline{\textbf{\unc{1.3}}}         & -           \\
& PSNR                   & -            & 19.2 \tightpm \unc{19.0}         & -           & -            & 11.0 \tightpm \unc{7.5}         & -            & -            & 7.7 \tightpm \unc{6.6}         & -           \\
& MI                     & -            & 4.5 \tightpm \unc{4.4}         & -           & -            & \underline{\textbf{5.8}} \tightpm \unc{2.5}         & -            & -            & 7.2 \tightpm \unc{3.0}          & -           \\
& Correlation            & -            & 2.7 \tightpm \underline{\textbf{\unc{0.6}}}       & -           & -            & 8.9 \tightpm \unc{1.4}         & -            & -            & 7.3 \tightpm \underline{\textbf{\unc{1.3}}}          & -           \\
\hline
& CV  & - & - & \underline{4.8} \tightpm \underline{\unc{1.2}} & - & - & \underline{4.3} \tightpm \underline{\unc{0.6}} & - & - & \underline{5.4} \tightpm \underline{\unc{0.7}}       \\
\hline

\end{tabular}}
\end{table}

\begin{table}[H]
\centering
\caption{Absolute errors (mean$\pm$standard deviation) of global SoS estimation (obtained with interpolation of metrics for SoS trial step of 25 m/s) are reported for three Simulation realizations and and six acquisitions per Phantom in three settings using Single, Dual, Full (17) frames as input to the methods, where applicable. 
Error values higher than 75\,m/s (50\% of SoS range tested) are highlighted in red. 
For each experimental setting (column), the lowest error (and similarly the standard deviation) in each group is highlighted in bold, with the lowest across all the groups also being underlined. 
Processing time required to calculate a metric at a single beamforming SoS is listed in the last column.}
\label{tab:combinedData32_interp25}
\resizebox{\textwidth}{!}{
\begin{tabular}{|@{\,}c@{\,}|l|ccc|ccc|ccc|c|}
\cline{2-12}
\multicolumn{1}{c|}{} & \multirow{2}{*}{Method} & \multicolumn{3}{c|}{Simulations} & \multicolumn{3}{c|}{Phantom 1} & \multicolumn{3}{c|}{Phantom 2} & \multicolumn{1}{c|}{Time} \\ 
\multicolumn{1}{c|}{} &                  & Single       & Dual         & Full        & Single       & Dual         & Full         & Single       & Dual         & Full        & [ms]   \\ 
\hline
\multirow{5}{*}{\rotatebox{90}{Quality}} & Focus                  & \textcolor{red}{80.6 \tightpm \unc{47.3}}      & 20.6 \tightpm \textbf{\unc{3.6}}         & \textbf{4.2} \tightpm \unc{3.7}        & \textcolor{red}{173.0 \tightpm \unc{19.9}}       & \textbf{12.9} \tightpm \textbf{\unc{8.0}}       & 9.0 \tightpm \textbf{\unc{1.4}}        & \textcolor{red}{132.4 \tightpm \unc{18.2}}       & 19.3 \tightpm \textbf{\unc{6.0}}       & 9.5 \tightpm \textbf{\unc{\unc{1.4}}}       & 16 \\
& Entropy                & \textcolor{red}{166.4 \tightpm \unc{62.7}}       & \textcolor{red}{124.8 \tightpm \unc{82.0}}       & 16.1 \tightpm \unc{5.0}        & \textcolor{red}{87.5 \tightpm \unc{46.2}}       & 18.7 \tightpm \unc{10.1}       & \textcolor{red}{77.8 \tightpm \unc{50.9}}       & \textcolor{red}{80.9 \tightpm \unc{26.8}}       & 62.7 \tightpm \unc{19.9}         & \textcolor{red}{98.5 \tightpm \unc{17.7}}        & 47 \\
& Tenengrad              & 41.5 \tightpm \unc{43.2}        & 23.3 \tightpm \unc{17.5}        & 28.2 \tightpm \unc{7.3}        & \textcolor{red}{160.0 \tightpm \unc{13.0}}       & \textcolor{red}{186.1 \tightpm \unc{5.3}}       & \textcolor{red}{179.9 \tightpm \unc{17.3}}       & \textcolor{red}{126.8 \tightpm \unc{15.5}}       & \textcolor{red}{159.0 \tightpm \unc{0.0}}       & \textcolor{red}{142.5 \tightpm \unc{19.2}}      & 0.76 \\
& ANACVF              & \textcolor{red}{140.9 \tightpm \unc{88.6}}          & \textcolor{red}{129.7 \tightpm \unc{85.2}}         & 42.6 \tightpm \unc{7.1}        & \underline{\textbf{64.5}} \tightpm \underline{\textbf{\unc{45.6}}}       & \textcolor{red}{108.3 \tightpm \unc{3.8}}       & \textcolor{red}{108.3 \tightpm \unc{3.8}}       & \textcolor{red}{102.3 \tightpm \unc{10.0}}       & \textcolor{red}{88.8 \tightpm \unc{6.8}}       & \textcolor{red}{102.8 \tightpm \unc{14.8}}       & 5.7 \\
& ST-Ten              & \underline{\textbf{3.8}} \tightpm \underline{\textbf{\unc{3.1}}}          & \textbf{9.0} \tightpm \unc{8.5}         & 10.5 \tightpm \textbf{\unc{2.5}}        & 106.5 \tightpm \unc{65.9}       & 13.0 \tightpm \unc{9.1}        & \textbf{6.6} \tightpm \unc{4.8}          & \textcolor{red}{88.3 \tightpm \unc{50.3}}       & \textbf{14.0} \tightpm \unc{9.0}         & \textbf{5.9} \tightpm 1.8        & 7.2 \\
\hline
\multirow{5}{*}{\rotatebox{90}{Comparison}} & SSIM                   & -            & 0.8 \tightpm \unc{0.4}         & -           & -            & 8.9 \tightpm \unc{2.5}         & -            & -            & \underline{\textbf{4.9}} \tightpm \unc{2.2}         & -           & 35 \\
& MSE                    & -            & 9.4 \tightpm \unc{1.0}         & -           & -            & \underline{\textbf{4.8}} \tightpm \underline{\textbf{\unc{0.6}}}          & -            & -            & 5.1 \tightpm \underline{\textbf{\unc{0.8}}}         & -           & 0.34 \\
& PSNR                   & -            & 9.5 \tightpm \unc{10.6}         & -           & -            & 16.9 \tightpm \unc{6.7}         & -            & -            & 10.8 \tightpm \unc{4.5}         & -           & 0.45 \\
& MI                     & -            & 6.4 \tightpm \unc{3.2}         & -           & -            & 8.6 \tightpm \unc{1.0}         & -            & -            & 10.2 \tightpm \unc{1.8}          & -           & 37 \\
& Correlation            & -            & \underline{\textbf{0.6}} \tightpm \underline{\textbf{\unc{0.3}}}       & -           & -            & 8.3 \tightpm \unc{0.7}         & -            & -            & 8.1 \tightpm \underline{\textbf{\unc{0.8}}}          & -           & 6.6 \\
\hline
& CV  & - & - & \underline{2.6} \tightpm \underline{\unc{2.2}} & - & - & \underline{4.2} \tightpm \underline{\unc{0.9}} & - & - & \underline{4.7} \tightpm \underline{\unc{0.8}}       & 62 \\
\hline
\end{tabular}}
\end{table}

\end{document}